\documentclass{jfm}
\usepackage{graphicx}
\usepackage{epstopdf, epsfig}

\shorttitle{Scaling laws for partially developed turbulence}
\shortauthor{Abigail Hsu, Ryan Kaufman, and James Glimm}

\title{Scaling laws for partially developed turbulence}
\author{Abigail Hsu\aff{1}
 \corresp{\email{abigail.hsu2009@gmail.com}},
 Ryan Kaufman\aff{1}
 \corresp{\email{rykauf@gmail.com}},
  \and James Glimm\aff{1}
  \corresp{\email{glimm@ams.sunysb.edu}}
  }

\affiliation{\aff{1}Stony Brook University, Stony Brook NY 11794}

\usepackage{setspace}
\usepackage{datetime}
\usepackage{epsfig}
\usepackage{graphicx}
\usepackage{latexsym,color,amsmath,amssymb}
\usepackage{enumerate}
\usepackage{amsfonts}
\usepackage{bbm}
\usepackage{hyperref}
\usepackage{cleveref}
\usepackage{caption}
\usepackage[all]{hypcap}
\usepackage{indentfirst}
\usepackage{float}
\usepackage{verbatim}
\usepackage{amsmath}
\usepackage{alltt}
\newcommand{\old}[1]{{}}
\def\hyph{-\penalty0\hskip0pt\relax}
\begin{document}
\pdfoutput=1

\maketitle

\begin{abstract}
We formulate multifractal models for velocity differences and gradients
which describe the full range of length scales in turbulent flow, namely: laminar, dissipation, inertial, and stirring ranges.
The models subsume existing models of inertial range turbulence. 
In the localized ranges of length scales in which the turbulence is only partially developed, we propose multifractal scaling laws with scaling exponents modified from their inertial range values. 
In local regions, even within a fully developed turbulent flow, the turbulence is not isotropic nor scale invariant due to the influence of larger turbulent structures (or their absence). For this reason, turbulence that is not fully developed is an important issue which inertial range study can not address.
In the ranges of partially developed turbulence,
the flow can be far from universal, so that standard inertial range turbulence 
scaling models become inapplicable. The model proposed here serves as a 
replacement.
Details of the  fitting of the parameters for the 
$\tau_p$ and $\zeta_p$ models in the dissipation range are discussed. Some of the behavior of $\zeta_p$ for larger $p$ is unexplained.
The theories are verified by comparing to high resolution simulation data.
\end{abstract}


\section{Introduction}
\label{sec:intro}

We develop a conceptual framework for
turbulent scaling laws across length scales extending beyond
the inertial range. 
Classically, fully developed turbulence is defined as occurring on length scales in which energy transfer is dominated by inertial forces. Partially developed turbulence is defined as turbulent (i.e. non-laminar) ranges in which turbulence may or may not be fully developed.
Thus, length scales with significant dissipation and stirring forces
are included within partially developed turbulence.
We propose new models with supporting verification which subsume and extend inertial and 
non-inertial range models of others. 
From our models and data analysis, we explain phenomena not previously observed and 
phenomena previously observed but not explained.

The inertial range is defined as an intermediate range of length scales, $l$, which are far from the Kolmogorov scale $\eta$ and integral scale $L$, i.e. $\eta \ll l \ll L$.
The dissipation range is the range of length scales between the laminar range and the inertial range. The stirring range, in many observational, experimental and simulation studies, is the range of length scales larger than the inertial range.  

Structure functions give a precise meaning to clustering of bursts of turbulent intensity and compound clustering (i.e. clustering of clusters) etc. They measure the dependence of these compound clustering rates on the length scale set by the observational size of the cluster. Two families of structure functions are studied here: one characterizes the powers of
the energy dissipation rate, $\epsilon$, and the other characterizes powers of
the velocity difference, $\delta u$.

The total turbulent energy dissipation rate $\epsilon$ in homogeneous turbulence is defined as the global spatial integral of:
\begin{equation}
\label {eq:eps}
\epsilon(\vec{x},t) = \frac{\nu}{2} \left(\frac{\partial u_i}{\partial x_j} + \frac{\partial u_j}{\partial x_i}\right)^2  \ ,
\end{equation}
where $\nu$ is the kinematic viscosity of the fluid and the summation convention is implied.
We define the coarse grained, local average of the
dissipation rate $\epsilon_l$ as,
\begin{equation}
\label{eq:avg_eps}
\epsilon_l(\vec{x}) = \frac{1}{|V_l|} \iiint_{V_l (\vec{x})}  {\epsilon(\vec{y},t)} ~ \mathrm{d} \vec{y} \ ,
\end{equation}
where $V_l (\vec{x})$ is a volume with a diameter $l$ centered around $\vec{x}$ in 3D. The coarse grained averaging means that $\epsilon_l$ reflects properties occurring on the length scale $l$.

The structure functions of $\delta_l u$ and $\epsilon_l$ satisfy
asymptotic scaling relations as a power of the length scale $l$.
The structure functions and the associated scaling exponents $\tau_p$ and $\zeta_p$ are given by the expectation relations,
\begin{equation}
\label{eq:scaling}
\langle {\left(\epsilon_l \right)}^p \rangle \sim l^{\tau_p}
\quad {\mathrm{and}} \quad
\langle |\delta_l u |^p \rangle \sim l^{\zeta_p} \ .
\end{equation}

The length scale dependent scaling exponents $\tau_p$ and $\zeta_p$ are obtained from logarithmic local slopes, i.e.
\begin{equation}
\label{eq:localslope}
\tau_p=\frac{d (\ln \langle {\epsilon_l}^p \rangle)}{d(\ln l)} 
\quad {\mathrm{and}} \quad
\zeta_p = \frac{d (\ln \langle {|\delta_l u|}^p \rangle)}{d (\ln l)}  \ ,
\end{equation}
as suggested by \cite{MilDim91}. 
\bigskip

The main thrust of the theory is that turbulent scaling laws remain in effect across all length scale ranges, but with modified length dependent scaling exponents.

A key step in the parameterization of the introduced scaling models is that $\tau_p$ and $\zeta_p$ in eq.~(\ref{eq:localslope}) are linear in $\ln l$ in the dissipation range.
Hence, the slopes of $\tau_p$ and $\zeta_p$ are constructed,
\begin{equation}
\label{eq:ZT_p}
T_p = \frac{d (\tau_p)}{d (\ln l)}  \quad {\mathrm{and}} \quad
Z_p = \frac{d (\zeta_p)}{d (\ln l)} \ ,
\end{equation}
which are modeled across all length scales as piecewise constant.

The piecewise constant values of $T_p$ and $Z_p$ appear to be a new discovery.
These piecewise constant values over the 4 length scale ranges are described as follows:
\begin{equation}\label{eq:Tp}
T_p = 
\begin{cases}
0, &\text{laminar range (LR)}\\
T^{dr}_p, &\text{dissipation range (DR)} \\
0, &\text{inertial range (IR)} \\
T^{sr}_p,&\text{stirring range (SR)}
\end{cases}
\qquad
Z_p = 
\begin{cases}
0, &\text{laminar range (LR)}\\
Z^{dr}_p, &\text{dissipation range (DR)} \\
0, &\text{inertial range (IR)} \\
Z^{sr}_p,&\text{stirring range (SR)}
\end{cases}
\end{equation}
In the inertial range, we take $T_p = 0$. We do not observe $T_p = 0$ from our data in this range, but include this to be consistent with the classical model by \cite{SheLev94}, denoted SL, in lieu of a more comprehensive theory. 

$T^{dr}_p$ and $Z^{dr}_p$ are constant in the dissipation range, and $T^{sr}_p$ and $Z^{sr}_p$ are constant in the stirring range for the problems we study.
These $T_p$ and $Z_p$ values are verified in the JHTDB and the $T_p$ values in the UMA data.
In addition, we observe that $T_p$ is linear in $p$, and $Z_p$ is given by an explicit $p$ dependent formula in the dissipation range. 

In the dissipation range, the logarithmic dissipation rate is proportional to $\ln (l)$. In contrast, the dissipation defined in the Navier-Stokes equation itself occurs in the laminar range at a rate proportional to the length scale $l$. For $p=2$, the laminar energy dissipation rate $\epsilon$ includes the classical viscosity $\nu$ within its definition. 
The increase of both $\tau_p$ and $\zeta_p$ for all $p$ in a limit as $l$ approaches $\eta$ from above
leads to negative constant values of $T_p$ and $Z_p$ in their respective dissipation ranges.

The key property of constant slopes $T_p$ and $Z_p$ is also satisfied in the stirring range and observed in the JHTDB data.
In principle, stirring forces can be added in an arbitrary manner at any length scale. 
Parameterization of the stirring range is not addressed in this paper. 

We provide a brief literature review in Sec.~\ref{sec:literature}. The numerical verification data from JHTDB and UMA are described in Sec.~\ref{sec:methods} as well as the methods used for data analysis. 
The scaling law results, which are the technical core of the paper,
are presented in Secs.~\ref{sec:taup_alllength} and~\ref{sec:zetap_alllength}.
The extended and refined scaling analytical methods
are discussed in Sec.~\ref{sec:CC}. 
A comment on the asymptotics of the viscous limit can be found in Sec.~\ref{sec:sl_conject}.
Conclusions are summarized in Sec.~\ref{sec:conc}.

\section{Inertial range prior results}
\label{sec:literature}
\cite{Kol41}, denoted K41, postulated universal laws to govern the statistics on
all such length scales in the inertial range in which the flow is statistically
self similar. Dimensional analysis in K41, based on the
self similarity hypothesis, led to the $-5/3$ scaling law.
However, because the energy dissipation for turbulent flows is intermittent, this model has been refined in various ways over the years to yield a multifractal scaling law. 
Summarized in \cite{Fri95}, the K41 exponent $\zeta_p$ is modified to capture the compound clustering of turbulent structure using a multifractal analysis. 
\cite{Kol62} refined his similarity hypothesis, denoted K62, added the influence of the large flow structure and included the influence of the intermittency.
The refined similarity hypothesis in \cite{Kol62} for the classical inertial range links the scaling exponent $\zeta_p$ of the longitudinal velocity structure and the scaling exponent $\tau_p$ of the energy dissipation rate as
\begin{equation}
\label{eq:dup_epsp}
\langle |\delta_l u |^p \rangle \sim \langle \epsilon_l^{p/3} \rangle l^{p/3} \ ,
\end{equation}
where $\langle {\epsilon_l}^{p/3} \rangle \sim  l^{\tau_{p/3}}$, or equivalently 
\begin{equation} 
\label{eq:sl_zeta_tau}
\zeta_p = \frac{p}{3} + \tau_\frac{p}{3} \ .
\end{equation}

The log-normal model from SL defines a theoretical model
for the PDF of the coarse-grained energy dissipation in the inertial range. SL studied the quantity $\epsilon^{(p)}_l$
defined as the ratio:
\begin{equation} 
\label{eq:succ_momt} 
{\epsilon_l}^{(p)} 
= \frac{\langle{\epsilon^{p+1}_l} \rangle}{\langle{\epsilon^{p}_l} \rangle} \ .
\end{equation}
where $p$ can be any non-negative integer.
The ${\epsilon^{(0)}_l} $ and $\epsilon^{(\infty)}_l$ are related to the mean fluctuation structure $\overline \epsilon$ and the filamentary structure.
The scaling law for $p \rightarrow \infty$ is ${\epsilon^{(\infty)}_l}  \sim l^{{-2}/{3}}$ .
As $p \rightarrow \infty$, the definition of $\tau_p$ stating that,
\begin{equation} 
\label{eq:tau_diffp}
\tau_{p+1} -\tau_{p} \longrightarrow -\frac{2}{3} \ ,
\end{equation}
or $\tau_{p} = -\frac{2}{3} \cdot p + C $.
The codimension $C$ is evaluated as $C = 3-1 = 2$ based on the assumption that the elementary filamentary structures have dimension $1$.
The expectation $\langle \epsilon_l^{p}\rangle$ has an $l$ dependence which is not
a pure exponential, but a mixture of exponentials, i.e. 
the scaling exponents for $\langle \epsilon_l^{p} \rangle$ are defined as
a weighted average of exponentials.

From the assumption of the interaction between structures of different order,
SL proposed the following relation between structures of adjacent order:
\begin{equation} 
\label{eq:tau_beta}
{\epsilon_l}^{(p+1)} = A_p {\epsilon_l}^{{\left(p\right)}^\beta} {\epsilon_l}^{{\left(\infty\right)}^{1-\beta}} \ .
\end{equation}
Based on eqs.~(\ref{eq:succ_momt}, \ref{eq:tau_beta}), SL derives a two step recursion for $\tau_p$. 
This recursion relation implies that $\tau_{p} = -\frac{2}{3} \cdot p + 2+ f(p)$,
where $f(\infty) = 0$ is assumed.
The equation for $f(p)$ has the solution $f(p) = \alpha \beta^{p}$ and with the boundary conditions $\tau_0 = \tau_1 = 0$, the solution becomes
\begin{equation} 
\label{eq:tau_sl}
\tau^{SL}(p) = -\frac{2}{3} \cdot p + 2\left[1 - \left(\frac{2}{3}\right)^{p}\right] \ .
\end{equation}
Substituting eq.~(\ref{eq:tau_sl}) into the relation shown in (\ref{eq:sl_zeta_tau}) yields
\begin{equation} 
\label{eq:sl_zeta}
\zeta_p = \frac{p}{9} + 2 \left[1 - \left(\frac{2}{3}\right)^{\frac{p}{3}}  \right] \ .
\end{equation}

\cite{Nov94} suggested that the -2/3 on the RHS of 
eq.~(\ref{eq:tau_diffp}) 
should be replaced by $-1$ based on the theory of infinitely
divisible distributions applied to the scaling of the locally averaged energy dissipation rate $\epsilon_l$. \cite{CheCao95}, denoted CC, accepted Novikov's suggestions and derived the formula:
\begin{equation} 
\label{eq:tau_cc}
\tau^{CC}(p) = -p + [(1+\tau_2)^{p} -1] /\tau_2 \ ,
\end{equation}
which uses the classical value $\tau_2  \approx -0.22$, which is derived from simulations, observations, experiments, and theory (SL) all in approximate agreement.

Kolmogorov proposed that $\zeta_3=1$ for incompressible, isotropic and homogeneous turbulence. \cite{FriDub95} showed that in the case of nonhomogeneous shell models with $\zeta_3 \neq 1$, the scaling of velocity structure functions in incompressible turbulence from SL still holds as $\zeta_p / \zeta_3 = p/9 +2[1 - (2/3)^{\frac{p}{3}}]$.

\cite{BolNor02} predicted a new scaling law for the scaling exponent $\zeta_p$ of velocity structure functions as
\begin{equation}
\label{eq:zetap_zeta3}
\zeta_p / \zeta_3 = p/9 +1 - (1/3)^{\frac{p}{3}} \ ,
\end{equation}
in supersonic turbulence for star formation based on a Kolmogorov\hyph{Burgers} model.
The same behavior is observed by \cite{MulBis00} in incompressible MHD.

The SL model has generated a considerable interest in the hierarchical nature of turbulence. Experiments and simulations have been conducted to evaluate the velocity and energy dissipation structures. 
\cite{ChaBau95_2,ChaBau95,ChaBau96} demonstrated experimental variables for the hierarchical structure assumption for the function $\zeta_p$ in eq.~(\ref{eq:sl_zeta}).
Experimental studies on a turbulent pipe flow and a turbulent mixing layer by ~\cite{ZouZhu03} verified the SL hierarchical symmetry. 
\cite{CaoCheShe96} showed agreement between the SL scaling exponents and high-resolution direct numerical simulations (DNS) of 3D Navier-Stokes turbulence.
Further references are added as needed through out the paper.

\section{Methods}
\label{sec:methods}

\subsection{JHTDB data}
\label{subsec:data}

We analyze the DNS data of the forced isotropic turbulence simulation from the Johns Hopkins
Turbulence Database (JHTDB) performed by \cite{LiPerWan08} and \cite{PerBurLi07}. 
The simulated flow has an integral scale Reynolds number $Re = 23,298$ and  a Taylor scale Reynolds number $Re_\lambda = 433$.  
The JHTDB data are generated by direct numerical simulation of forced
isotropic turbulence in a cubic domain with length $L = 2\pi$ and
periodic boundary conditions in each direction. The simulation has a
resolution of $1024^3$ of cells.
Energy is injected to maintain a constant value for the total energy. The JHTDB data are collected after the simulation has reached a statistical stationary state. The data are posted on the website
\centerline{\url{http://turbulence.pha.jhu.edu}}
The JHTDB data focus on analysis of the inertial range.
Because of this emphasis, its coverage of the dissipation and stirring ranges is limited.  
The ratio of the Kolmogorov length scale to the computational grid space is $\eta / \Delta x = 0.46$.
Thus, the JHTDB data do not fully resolve the Kolmogorov length scale $\eta$. 
With these data, we confirm many aspects of our scaling law model.
We anticipate the need for additional simulation data 
such as the $4096^3$ cell data from the JHTDB
in further analysis of laminar and dissipation ranges.

\subsection{JHTDB data analysis}
\label{subsubsec:analysis_methods}
\begin{figure}
\center{\includegraphics[width=1.\textwidth]{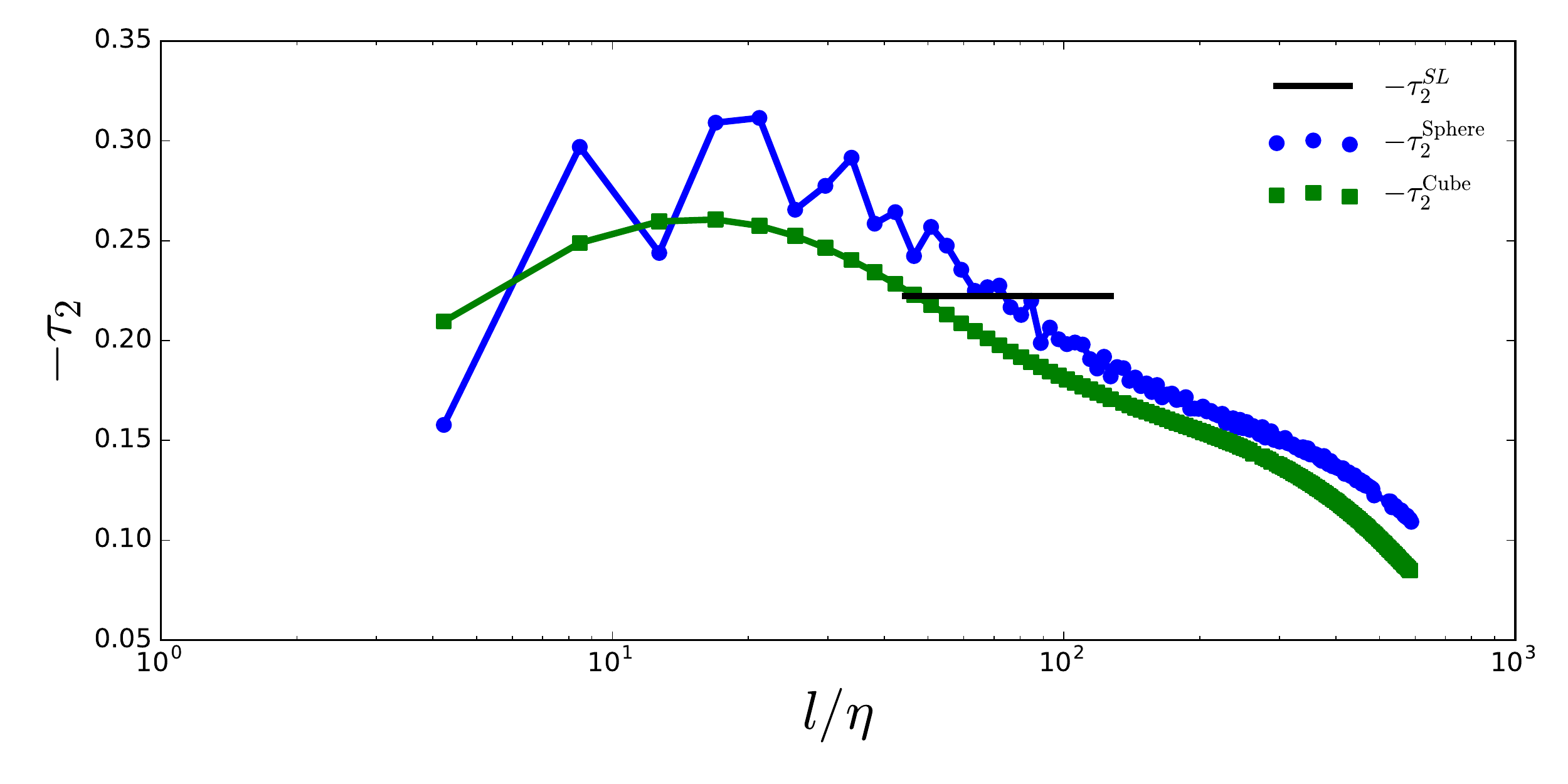}}
\caption{\label{fig:compare_ball_cube}
Comparison of the cubic to the spherical 
average with a diameter $l$ for $\epsilon_l$ in the computation of $-\tau_2$.  The horizontal solid line is the constant local exponent $-\tau^{SL}_2$ value from SL and is located at the inertial range that is determined by $\zeta_2$ from JHTDB.}
\end{figure}

The velocity differences have a tensorial dependence on the velocity component directions and the differencing directions. 
The longitudinal direction is more convenient for experiments, and many experimental prior studies focused on the longitudinal velocity increment based on Taylor's hypothesis. Details are described by \cite{HeChe99}. 
We define the longitudinal velocity increment as
${\delta_l} u(x,y,z,t) = u \left(x + l,y,z, t \right) - u(x,y,z, t)$,
where $u$ is the $x$ component of velocity.

The coarse-graining length scale, $l$, in the structure function definition, eq.~(\ref{eq:avg_eps}) is implemented by a 3D average over distances of a scale $l$, and $\epsilon$ is given at discrete locations in space.
From previous literature, the averaging
volume is taken to be a sphere or a cube. The definition of $l$ as radius or diameter is not consistent in the literature.
In this paper, we distinguish between spherical and cubic average with a diameter $l$.
The dominant effects on the averaging element, sphere or cube, come from the most extreme points on the elements, i.e. the boundary of the sphere or corners of the cube, and always with these extreme points at distance $l/2$ from the
center.

In Fig.~\ref{fig:compare_ball_cube}, we observe an approximate but not exact agreement between the cubic and spherical averages. 
The cubic average, producing a consistent but less noisy range, is used for the local average of the energy dissipation rate in this paper.

A further observation is that $\tau_2$ in the inertial range is not consistent with prevailing theory i.e. $\tau_2$ is not constant while $\zeta_2$ is constant in this range.
Fig.~\ref{fig:compare_ball_cube} shows the theoretical value for the inertial range $\tau^{SL}_2$ as a horizontal line. The local slope measurements are points connected by line segments. While an average across a large range is consistent with SL, no universal constant local slope is found that is consistent with $\tau^{SL}_2$.
For this reason, we regard the $\tau_p$ theory from SL more exactly as a theory for $\zeta_p$ rather than a theory for $\tau_p$.

\subsection{UMA data analysis}
\label{subsec:uma_data}
\begin{figure}
\center{\includegraphics[width=1.\textwidth]{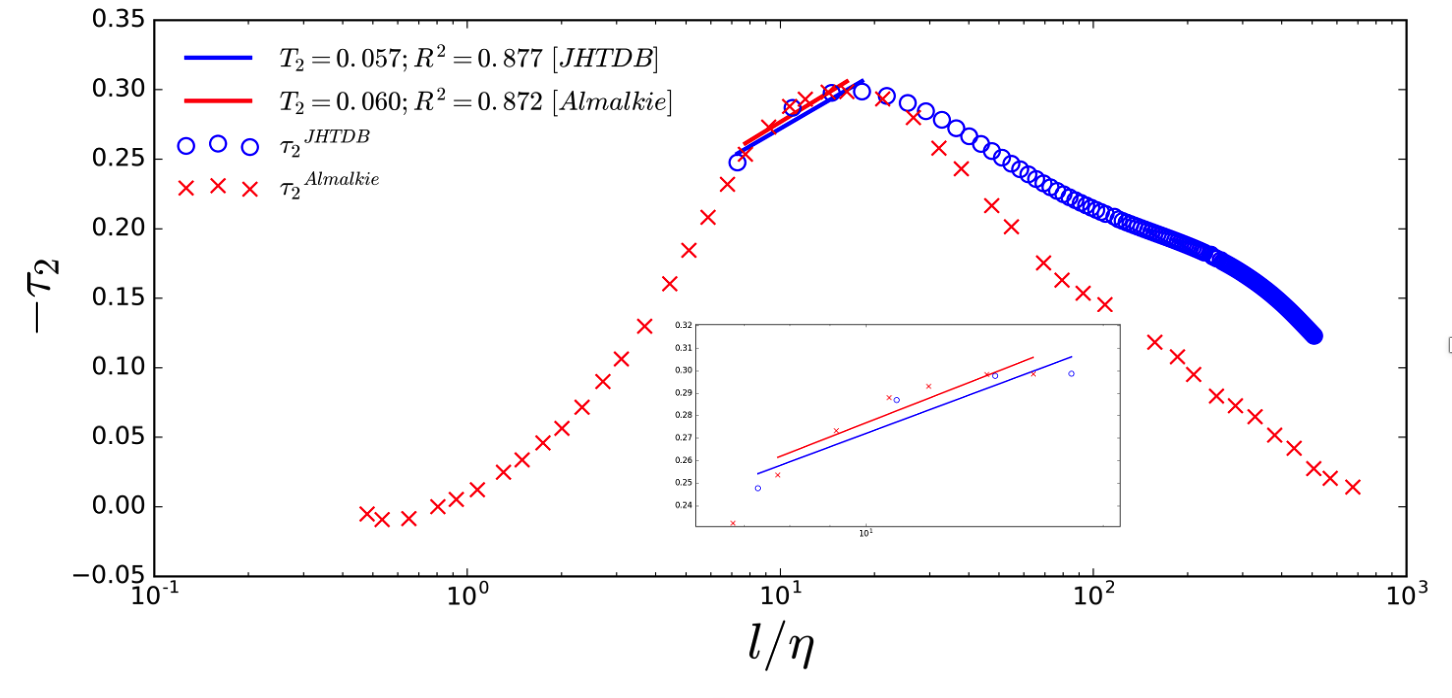}}
\caption{\label{fig:tau2_JHAlm} Consistency of the JHTDB and UMA data. The inset is a refined image of the overlapped dissipative ranges.}
\end{figure}

The UMA data from \cite{AlmSte12} are based on a DNS of isotropic homogeneous turbulence from the University of Massachusetts Amherst (UMA). This simulation uses a third-order Adams-Bashforth and pseudo-spectral method. 
The simulation uses a periodic cube with edge length $2 \pi$ and a $2048^3$  numerical grid. 
The simulation parameters are $\eta / \Delta x= 3.04$ and $\eta k_{max} = 6.4$.
The flow has an integral scale Reynolds number $Re = 3,426$ and  a Taylor scale Reynolds number $Re_\lambda = 151$.  
The inertial range turbulence is fully developed and the data also describe with a complete dissipation range with length scales smaller than the Kolmogorov length scale.
There are sufficient data to provide verification in the laminar range.
We digitized the local slopes $-\tau_p$ from the published UMA data for $p=2,3,4$ by \cite{AlmSte12}.

The UMA locally averaged dissipation rate $\epsilon_l$ data are defined as a spherical average with a diameter $l$. An interpolation consistent with the numerical method is used to solve the governing equations and allows the elimination of the noise.

As noted in Fig.~\ref{fig:compare_ball_cube}, the spherical average leads to a higher $-\tau_p$ value than the cubic average. A vertical shift of 0.04 for $-\tau_2$ of the cubic averaged JHTDB data is needed to reach the spherical averaged UMA maximum for comparison. 
In addition, a horizontal shift of the normalized length scale $l/\eta$ is needed to compensate for the JHTDB under resolution of the Kolmogorov scale. 
In Fig.~\ref{fig:tau2_JHAlm}, we see that the linear local slope is clearly defined in the dissipation range from the UMA $-\tau_2$ data and extends the JHTDB data. A refined inset shows the overlap from the two datasets after these shifts.

\subsection{Schematic model formulation}
\label{subsec:schemati}
\begin{figure}
\center{\includegraphics[width=1.\textwidth]{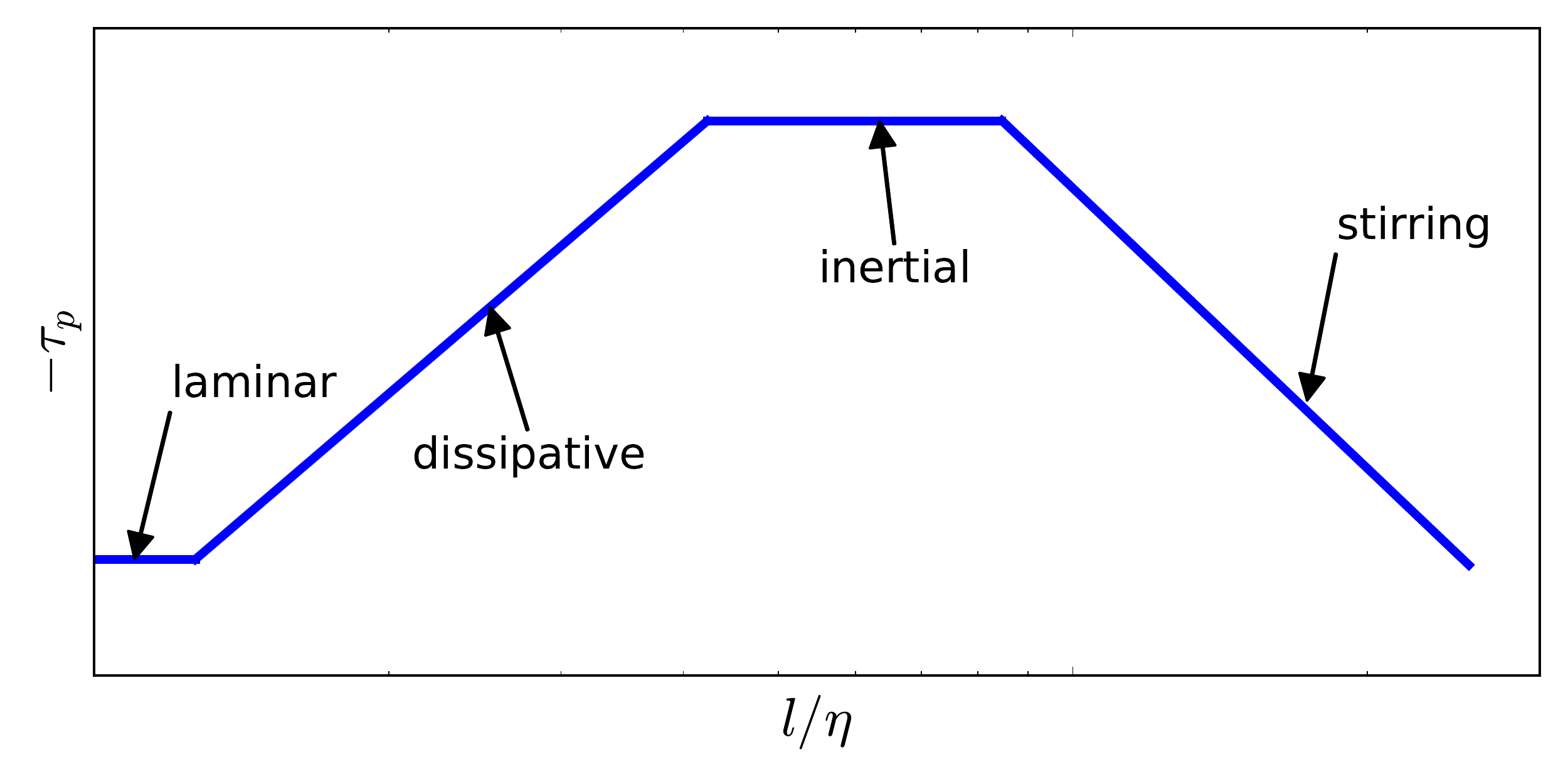}}
\caption{\label{fig:idea_tau2} Schematic representation of $-\tau_p$ vs. $l/\eta$ on a semi-log scale.}
\end{figure}

Fig.~\ref{fig:idea_tau2} summarizes our major ideas. 
The laminar, dissipation, inertial and stirring ranges are labeled. 
This figure displays the slopes, $T_p$, as linear segments in the dissipation and stirring ranges. 
In Figs.~\ref{fig:compare_ball_cube} and~\ref{fig:tau2_JHAlm}, $-\tau_2$ does not show a clear flat range to indicate the classical inertial range. 
The flat segment shown in the schematic representation is taken from the SL theory. We comment on its absence from the JHTDB and UMA data in Sec.~\ref{subsec:full_taup}.

\begin{figure}
\begin{minipage}[b]{.48\textwidth}
	\centering
   \includegraphics[width=1.0\textwidth]{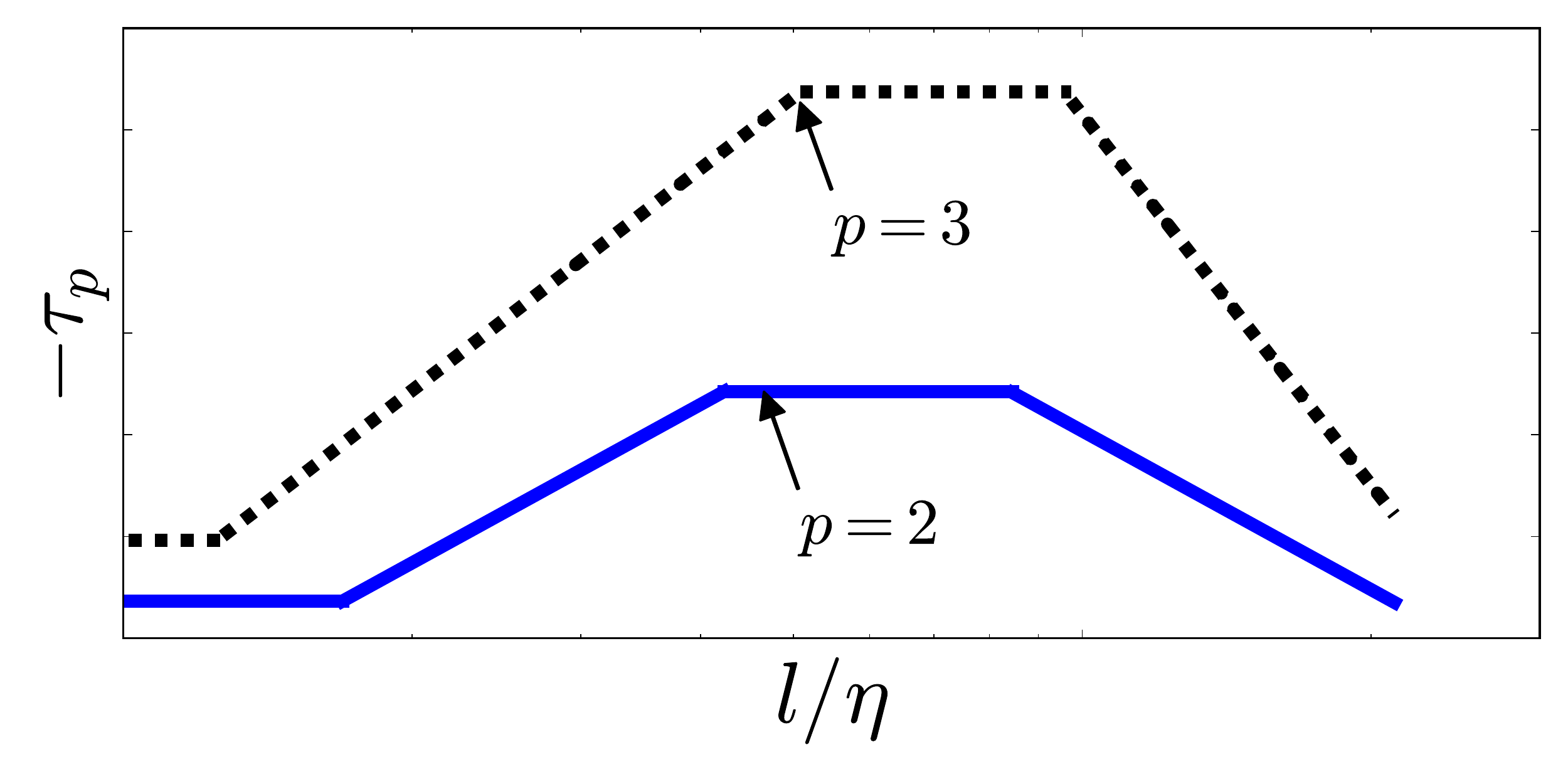}
 	\caption{
Schematic representation of the local slope, $-\tau_p$
vs. log length scale with the $p$-dependence shown by the arrow.
}
   \label{fig:idea_taup}
\end{minipage}\hfill
\begin{minipage}[b]{.48\textwidth}
   \includegraphics[width=1.0\textwidth]{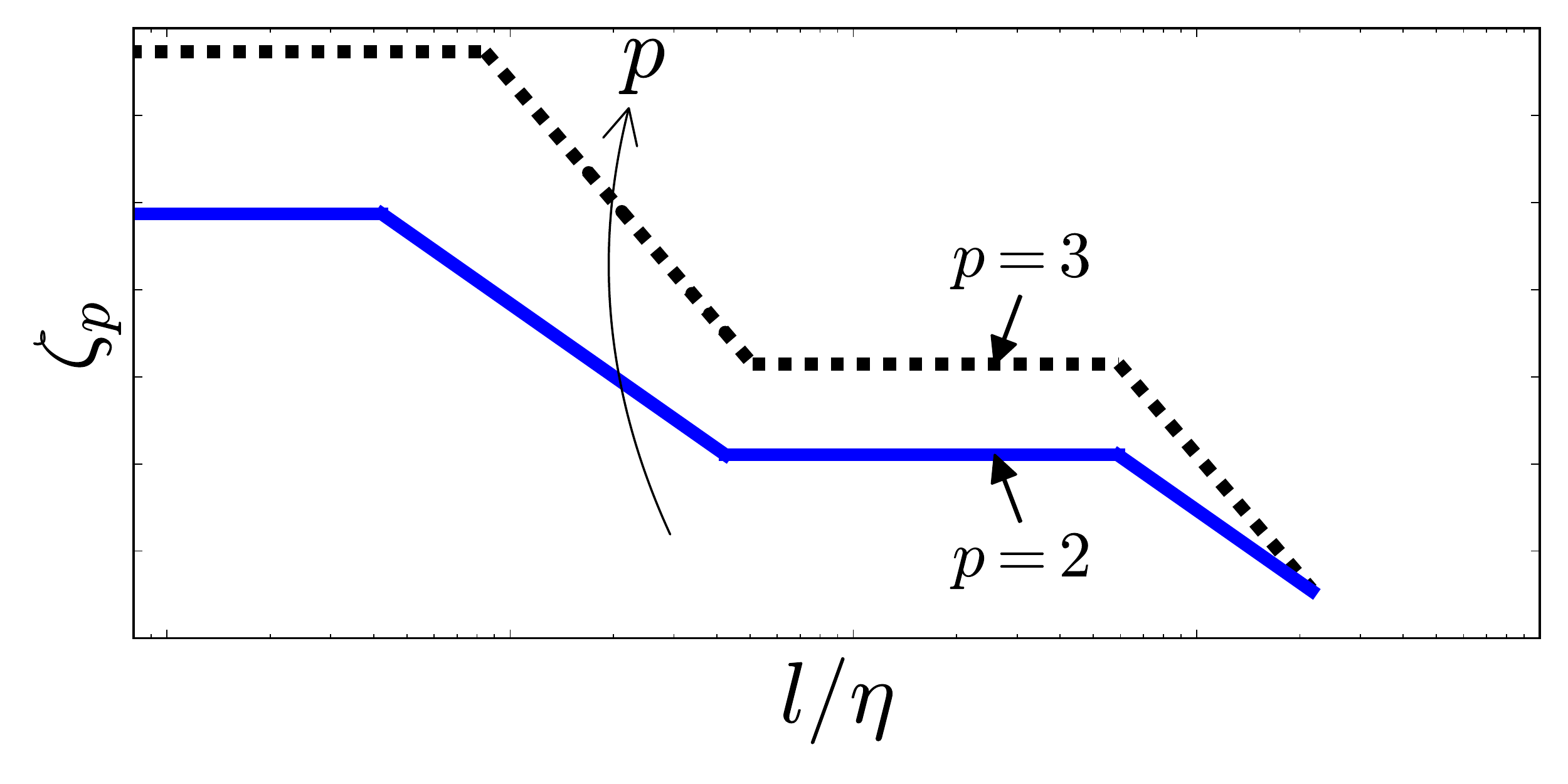}
	\caption{
Schematic representation of the local slope, $\zeta_p$ vs. log length scale with the $p$-dependence shown by the arrow.
}
   \label{fig:idea_zetap}
\end{minipage}
\end{figure}

Figs.~\ref{fig:idea_taup} and~\ref{fig:idea_zetap} show that as $p$ increases, so do the values for $-\tau_p$ and $\zeta_p$.
Laminar flow occurs at a nearly zero and is 
shown as a nearly horizontal line 
on the semi-log plots in Figs.~\ref{fig:idea_tau2}, ~\ref{fig:idea_taup} and~\ref{fig:idea_zetap}. The linearity of $\tau_p$ in the dissipation range implies the equation 
\begin{equation}
\label{eq:taup_1}
\tau_p = T_p \ln \left(\frac{l}{\eta} \right) + b_p \ ,
\end{equation}
where $T_p$ represents the constant slope observed in $\tau_p$ and $b_p$ is the $\tau_p$ value at the Kolmogorov length scale ($\eta$). 

Similarly, in the dissipation range, we define
\begin{equation}
\label{eq:zeta_linear}
\zeta_p = Z_p \cdot \ln \left(\frac{l}{\eta} \right) + a_p \ ,
\end{equation}
where $Z_p$ represents the constant slope observed in $\zeta_p$ and $a_p = \zeta_p$ at the Kolmogorov length scale ($\eta$).

A transition point in the schematic representation occurs at the length scale where the slope of $\tau_p$ or $\zeta_p$ changes.
The transition from turbulent dissipation to laminar dissipation 
occurs approximately at $l = \eta$. The detailed locations of the transition are dependent 
on $p$ and are different for $\zeta_p$ and $\tau_p$. 
The transition actually occurs gradually rather than discontinuously as we see in Fig.~\ref{fig:laminor_tau234_JHAlm}. 
However, in our modeling, $T_p$ and $Z_p$ are piecewise constant in $\ln l$ as an approximation to an exact theory. 
All other transitions are determined similarly.

In this modeling, $\zeta_p$ or $\tau_p$ for each $p$
defines its own inertial range. 
The SL model holds within the inertial range defined by $\zeta_p$.

\section{Scaling laws for the energy dissipation rate}
\label{sec:taup_alllength}

\subsection{The laminar range}
\label{subsec:Dis_Kol}
\begin{figure}
\center{\includegraphics[width=1.\textwidth]{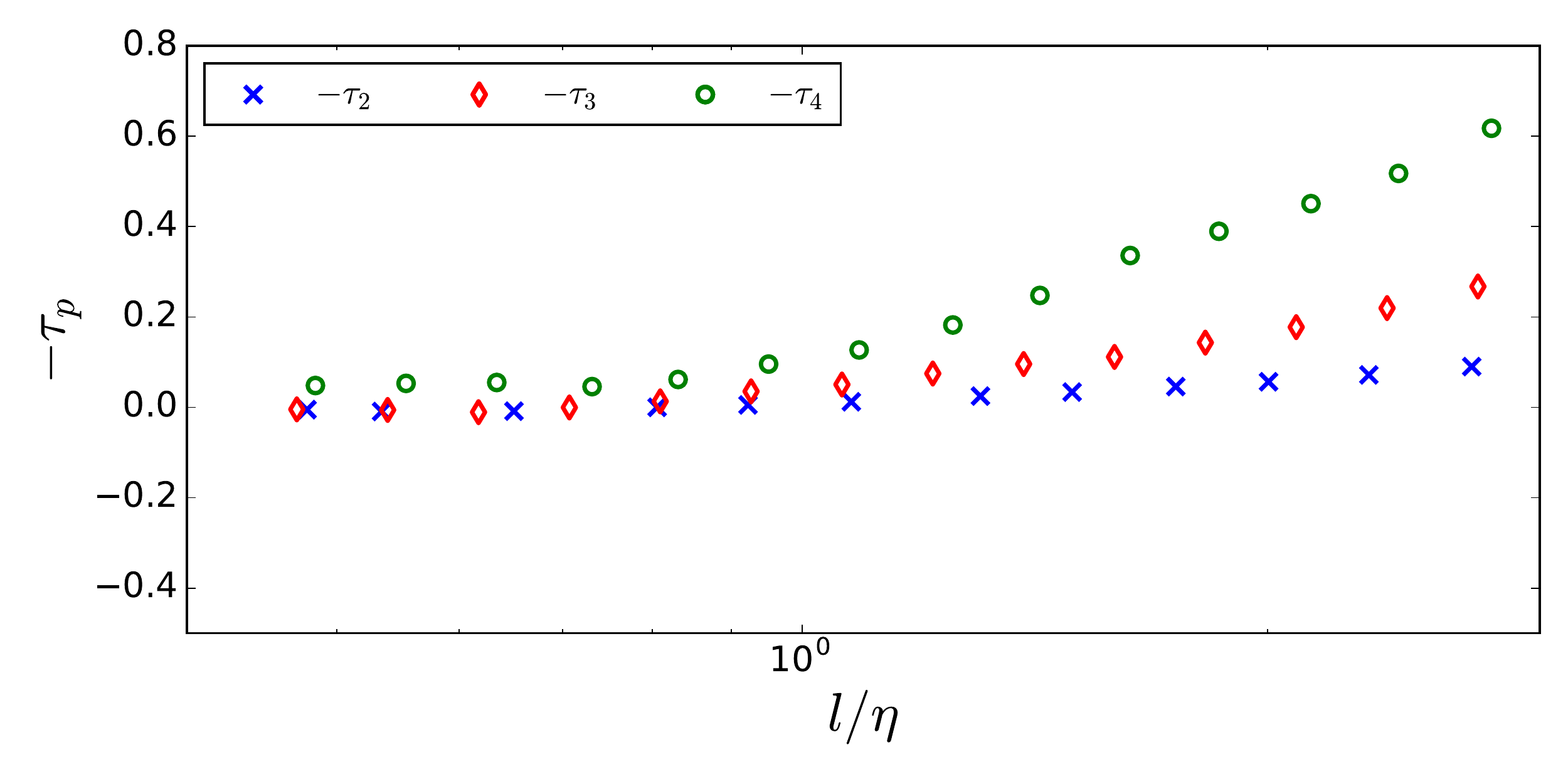}}
\caption{\label{fig:laminor_tau234_JHAlm} The digitized $-\tau_p$ values from the UMA data enlarged for the laminar range, for $p=2,3,4$. We observe constant $-\tau_p$ values with weak $p$ dependence for smaller lengths.}
\end{figure}

The digitized UMA data of $-\tau_p$ for $p=2,3,4$ in the laminar range is shown in Fig.~\ref{fig:laminor_tau234_JHAlm}.
$-\tau_2$ and $-\tau_3$ are close to zero here with, perhaps, a small $p$ dependence shown in the $\tau_4$ data.
Thus, we assume that the $\tau_p$ values are approximately zero in the laminar range in which $l/\eta < 1$.

\subsection{The $p$-dependent linear slopes in the dissipation range}
\label{subsec:Dis_linear}

\begin{figure}
\center{\includegraphics[width=1.\textwidth]{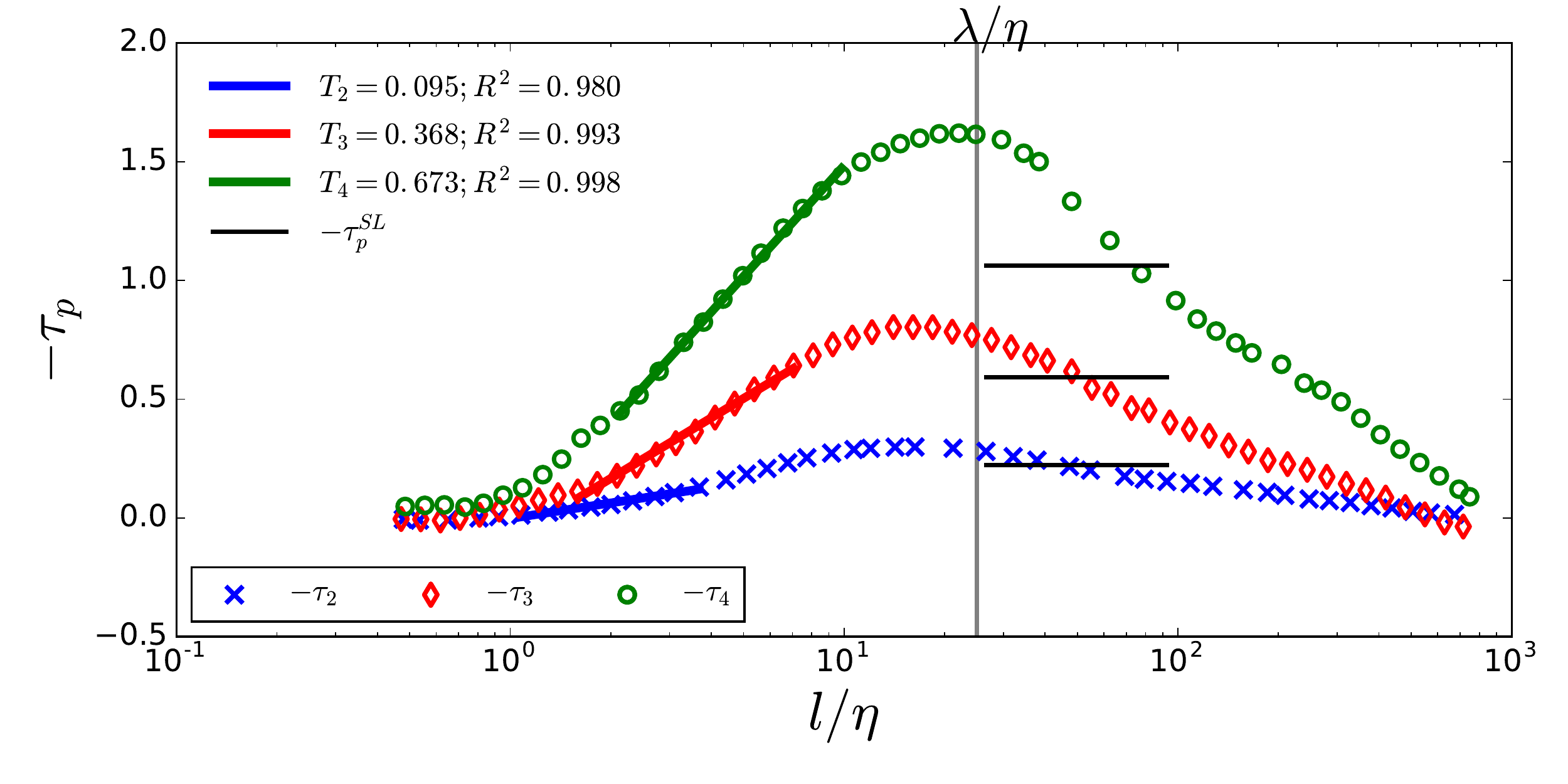}}
\caption{\label{fig:tau234_JHAlm} The digitized $-\tau_p$ UMA data. The horizontal solid lines are the constant local exponent $-\tau_p$ values from SL with the inertial range determined by $\zeta_2$ from JHTDB. The vertical line marks the Taylor micro scale.}
\end{figure}

\begin{figure}
\centering
\begin{minipage}{.48\textwidth}
	\centering
   \includegraphics[width=1.0\textwidth]{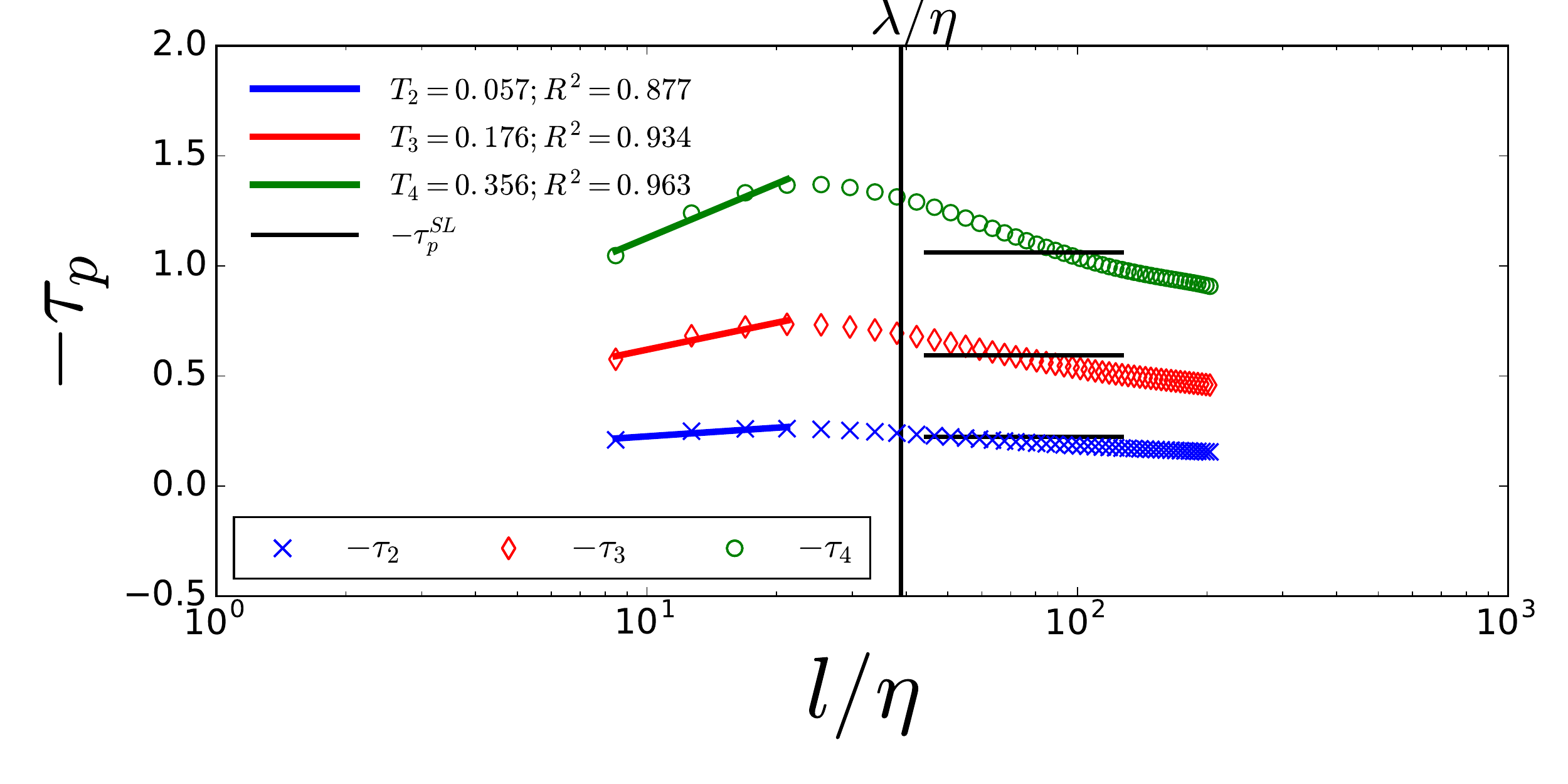}
   \vfill
 	\caption{
		The local slopes $-\tau_2, -\tau_3, -\tau_4$ from the JHTDB data are linear in $\ln (l/\eta)$ in the dissipative ranges, with constant slopes $T_p$. An $R^2$ value closes to $1$ indicates the goodness of the fit.
}
   \label{fig:SLCCtau234_jh}
   \vfill
\end{minipage}
\hfill
\begin{minipage}{.48\textwidth}
	\vskip 4.5mm
   \includegraphics[width=1.0\textwidth]{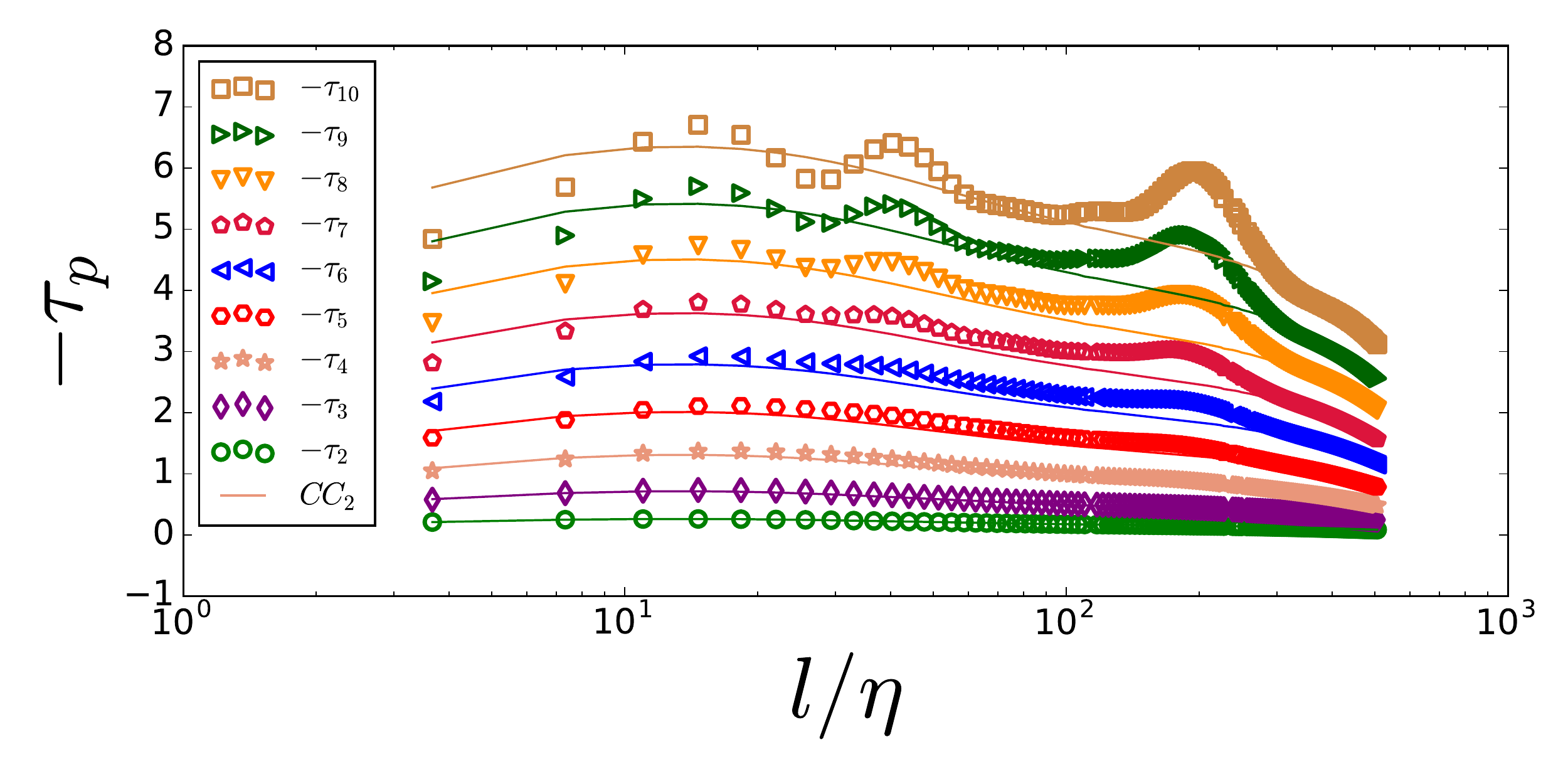}
	\caption{
		The local slope $\tau_p$ for all $2 \leq p \leq 10$ with $p$ increasing from the bottom to the top. The JHTDB data are shown in scatter points and the modeled $-\tau_p$ values are in solid lines from the $CC_2$ model explained in Sec.~\ref{sec:CC}.
}
   \label{fig:E_tau_allp_fit}
\end{minipage}
\end{figure}

The UMA data of Fig.~\ref{fig:tau234_JHAlm} and the JHTDB of Fig.~\ref{fig:SLCCtau234_jh} show the linear segment of $\tau_p$ in the dissipation range.
With limited data in small length scales but including higher $p$ values (up to
$p = 10$), we reach the same conclusion from 
Fig.~\ref{fig:E_tau_allp_fit} for the JHTDB data. The solid lines are modeled by the $CC_2$ model and will be explained in Sec.~\ref{sec:CC}. 
The linear slopes are determined by least squares.

\begin{figure}
\center{\includegraphics[width=1.\textwidth]{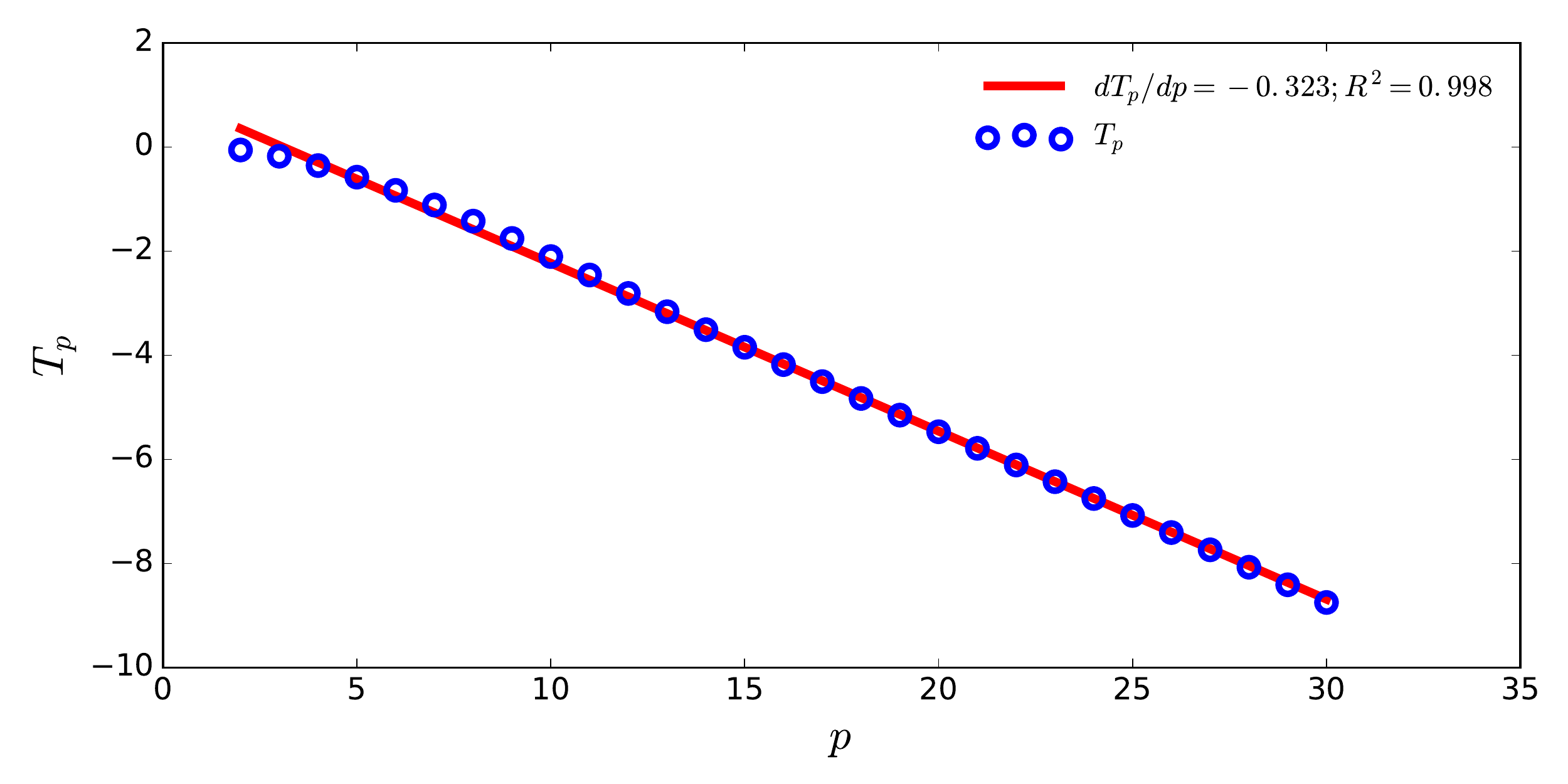}}
\caption{\label{fig:T_p} The $T_p$ in the dissipative range is linear for $p$ in the range $2 \leq p \leq 30$ for the JHTDB data.}
\end{figure}

Recall that $\tau_p = T_p \ln \left(l/\eta\right) + b_p$. In Fig.~\ref{fig:T_p}, we plot the $T_p$ for $p$ up to 30 in the dissipation range. The slopes, $T_p$, are linear in $p$, with a $T_2$ data dependence, i.e.
\begin{equation}
\label{eq:Tp}
T_p = \frac{d T_p}{d p}  \cdot  p + (T_2 - 2 \cdot \frac{d T_p}{d p}) \ .
\end{equation}

We find that ${d T_p}/{d p} = -0.323$ from the JHTDB data.

\subsection{Full \text{\Large$\tau_p$} parameterization}
\label{subsec:full_taup}

We have developed a model that captures all of the length scales from the laminar range up to the small length scale
end of inertial range for the energy dissipation rate 
Based on eq.~(\ref{eq:Tp}) and the assumption of $\tau_p \approx 0$ in the laminar range.
A model for $\tau_p$ in the dissipation range is
\begin{equation}
\label{eq:full_parm_taup}
\tau^{dr}_p = \left(\frac{d T_p}{d p}  \cdot  p + \left(T_2 - 2 \cdot \frac{d T_p}{d p}\right) \right)  \cdot  \ln \left(\frac{l}{\eta} \right)  \ ,
\end{equation}
which can be extended for length $l$ up to the Taylor micro-scale $\lambda$.

In the JHTDB and UMA data, there is no flat inertial range observed for $\tau_p$ and the $\tau_p$ peak occurs approximately at the Taylor micro-scale $\lambda$. The peak could be an isolated point represents a transition to the stirring range. Alternatively, there maybe a small inertial range and a large transition range, in which case, it is possible that with higher Reynolds number, a true (flat) inertial range for $\tau_p$ may appear.

\section{The extended and refined CC models for the scaling exponent of the energy dissipation rate}
\label{sec:CC}
\begin{figure}
\center{\includegraphics[width=1.\textwidth]{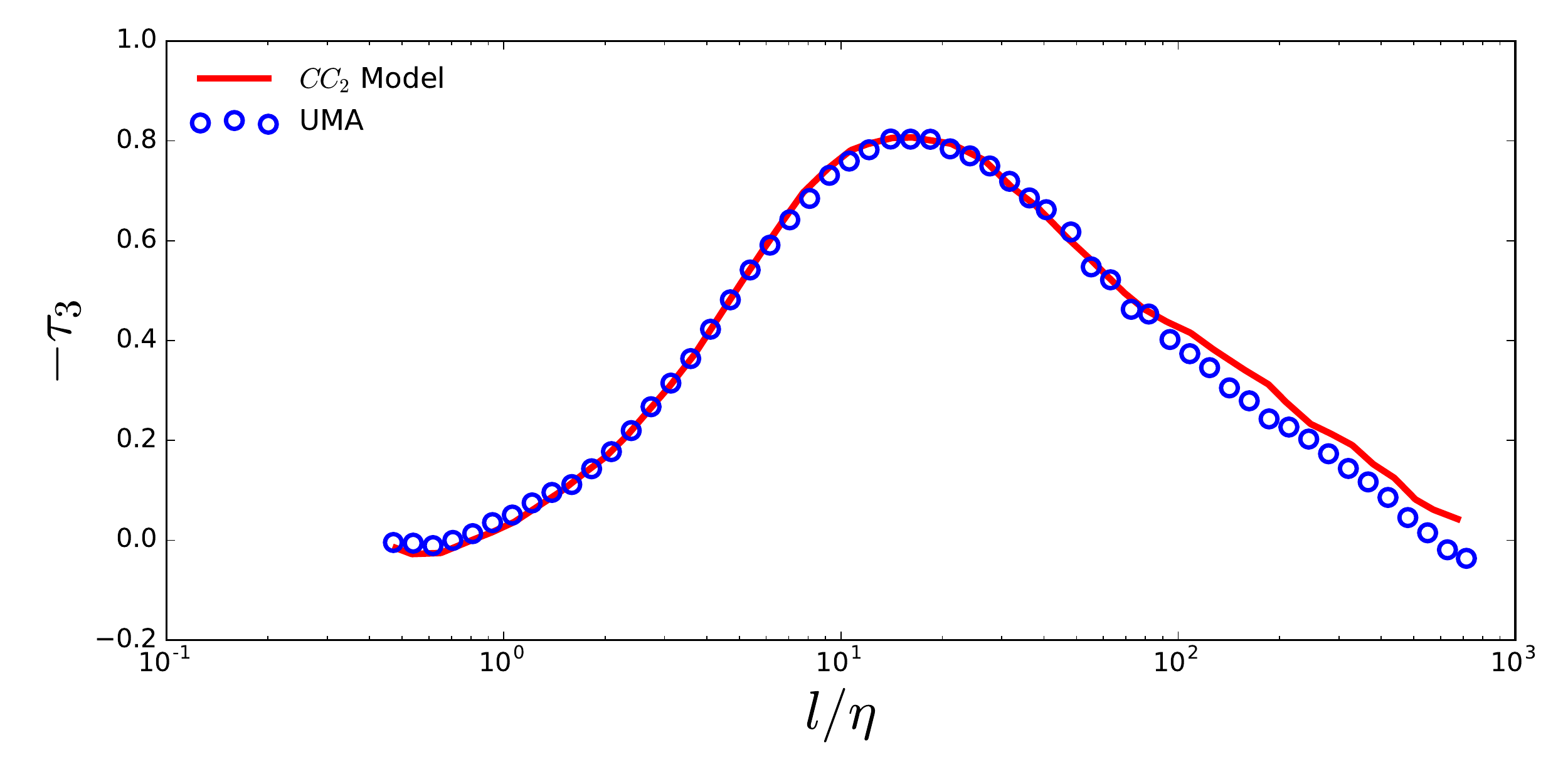}}
\caption{\label{fig:tau3_CC}
The extended $CC_2$ model for $-\tau_3$ shows agreement with the UMA data across all measurable length scales.}
\end{figure}

The CC model defined by \cite{CheCao95} shown in eq.~(\ref{eq:tau_cc}) has a dependence on inertial constant $\tau_2 = {\tau^{SL}_2} \approx -0.22$ value from SL. 
We extend the model beyond the inertial range, where a constant $\tau^{SL}_2$ is no longer an accurate value for length scale dependent $\tau_2$. 
To calculate $\tau_p$ for $p > 2$, we substitute the $\tau_2$ data that are measured across all length scales into eq.~(\ref{eq:tau_cc}).
The extended model is denoted as the $CC_2$ model.
Fig.~\ref{fig:tau3_CC} shows this model for $\tau_3$ (solid line) in comparison to the observed $\tau_3$ value (data points).
The same comparison for $p \leq 10$ is shown in Fig.~\ref{fig:E_tau_allp_fit}.
The data and model values agree for small $p \leq 5$, but diverge at higher $p$ values.
\begin{figure}
\center{\includegraphics[width=1.\textwidth]{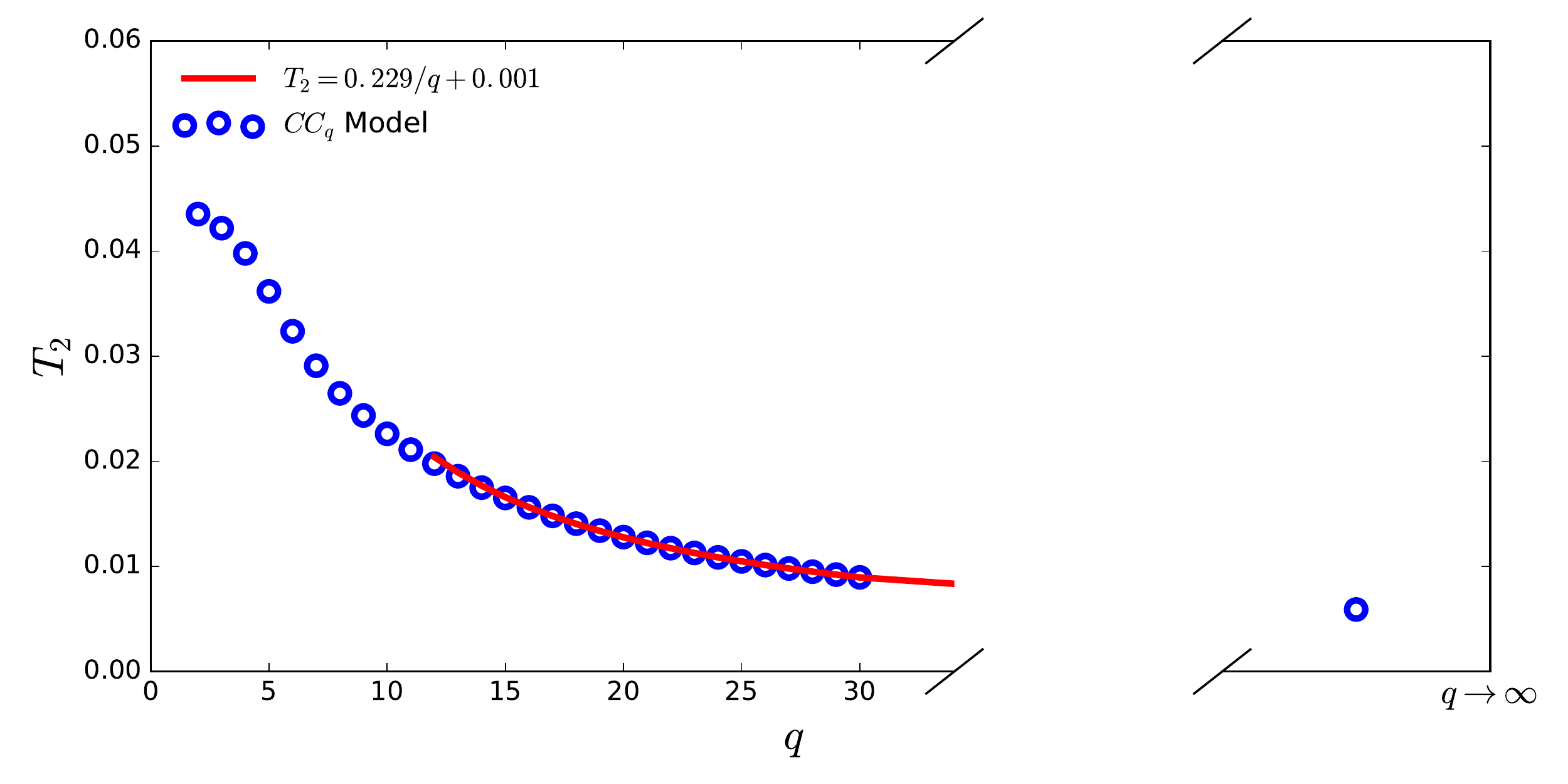}}
\caption{\label{fig:tau2_CCp}The linear slopes $T_2$ of $\tau_2$ calculated from the $CC_q$ model for $q$ from $2$ to $30$ using JHTDB is plotted vs. $q$ in the dissipative range.}
\end{figure}
We propose a modification of $CC_2$, which we denote as $CC_q$ model. 
Instead of taking a measured $\tau_2$, as in our extended $CC_2$ model, we substitute the measured $\tau_q$ data that is available across all measurable length scales, and solve numerically for $\widehat{\tau}_2$ shown as below: 
\begin{equation}
\label{eq:cc_q}
\tau_q = -q + [(1+\widehat{\tau}_2)^{q} -1] /\widehat{\tau}_2 \ .
\end{equation}
Then, the calculated $\widehat{\tau}_2$ can be substituted back to eq.~(\ref{eq:tau_cc}) to calculate $\tau_p$ for any $p$.
From this point, all $\tau_p$ values can be modeled as a function of $p$ and the measured $\tau_q$ using the $CC_q$ model.

Fig.~\ref{fig:tau2_CCp} shows the linear fit $T_2$ of $\tau_2$ that is calculated from the $CC_q$ model using measured $\tau_q$ values in the dissipation range, for $q=2,3,...,\infty$.
This sequence of $T_2$ values are fitted by:
\begin{equation}
T_2 = 0.229/q+0.001 
\end{equation}
for any $q$ value. 
$T_2$ decreases and converges asymptomatically to $0.001$ as $q \to \infty$. Hence, the $CC_\infty$ model is defined as $CC_\infty = \lim_{q \to \infty} CC_q$. 

\section{Scaling laws for the longitudinal velocity increment structure functions}
\label{sec:zetap_alllength}

\begin{figure}
\center{\includegraphics[width=1.\textwidth]{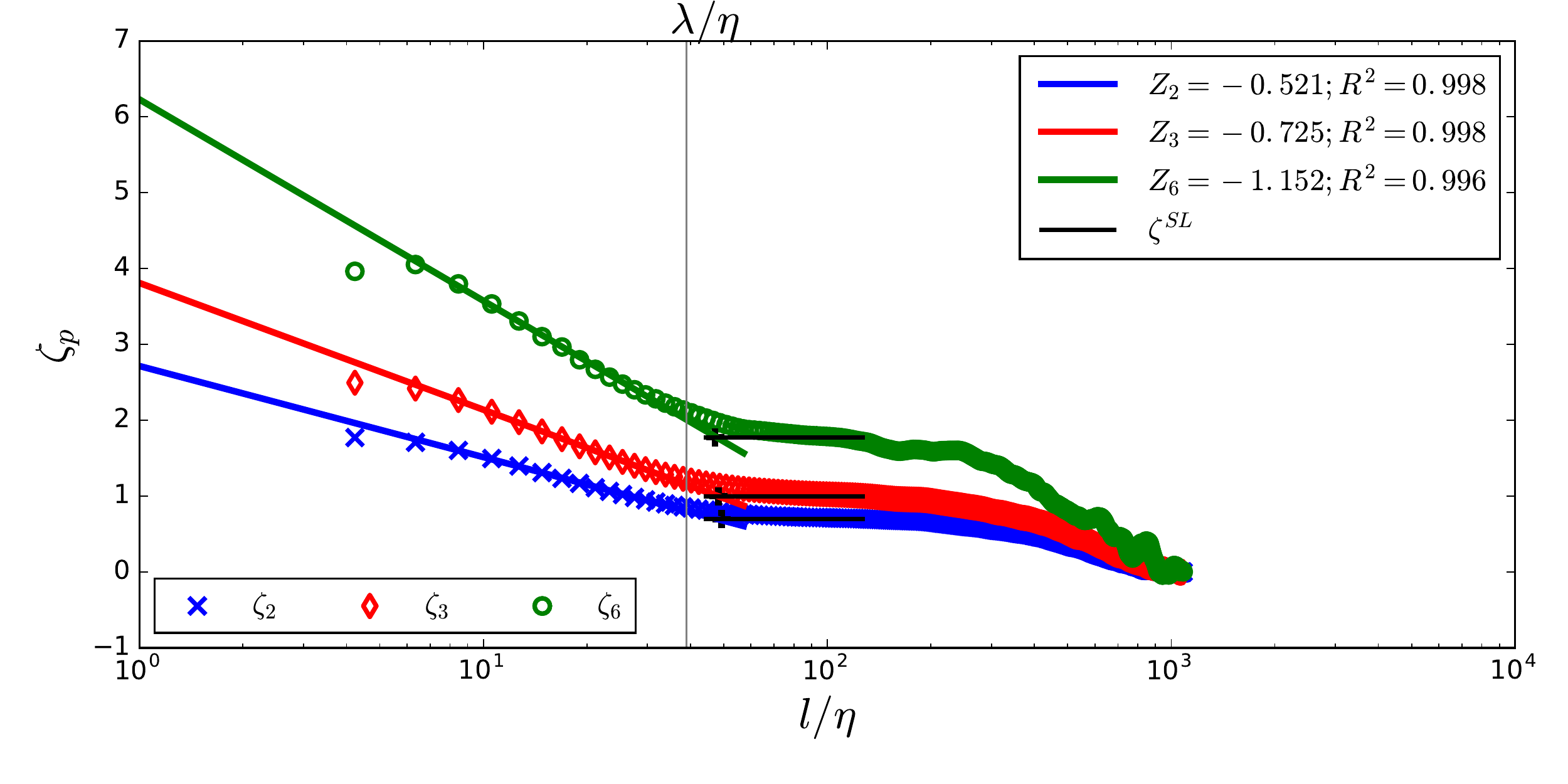}}
\caption{\label{fig:zeta236_fit} $\zeta_2, \zeta_3, \zeta_6$ are linear in the dissipative ranges with the solid line representing the linear fit of $\zeta_p$ in the dissipation range, extrapolated to the Kolmogorov scale for the JHTDB data. The horizontal lines are the corresponding SL model values  and the vertical line marks the Taylor micro-scale. The bold `+' near the Taylor scale values $\lambda/\eta$ are the transition points from the dissipation ranges to the inertial ranges.}
\end{figure}

\begin{figure}
\center{\includegraphics[width=1.\textwidth]{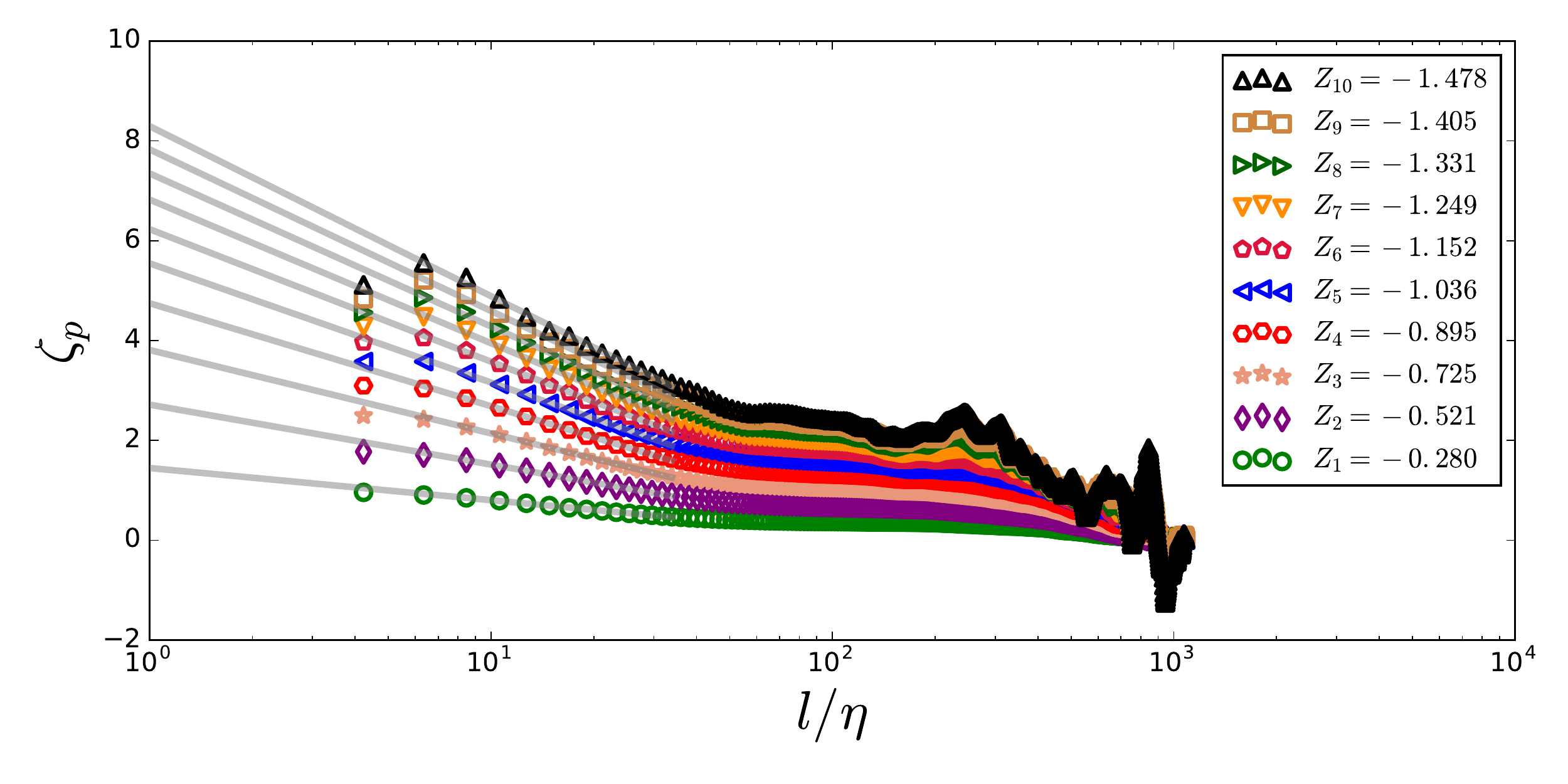}}
\caption{\label{fig:L_zeta_allp_fit} The JHTDB defined $\zeta_p$ with $p$ increasing from $1$ to $10$ from the bottom to the top across all available length scales. The plot shows scatter points with solid lines representing the linear fit of $\zeta_p$ in the dissipation range extrapolated to the Kolmogorov scale.}
\end{figure}

\subsection{The $p$-dependent linear slopes in the dissipation range}
\label{subsec:zeta_diss}

The JHTDB data for the slopes $Z_p$ of the longitudinal velocity increment $\langle |\delta_l u_1|^p \rangle$
vs. $\ln (l/\eta)$ for $p = 2,3,6$ are shown in Fig.~\ref{fig:zeta236_fit}.
We see that $\zeta_p$ is consistent with the horizontal solid lines of $\zeta_p^{SL}$ values obtained from eq.~(\ref{eq:sl_zeta}) in the inertial range.
In addition, $\zeta_p$ is linear in $\ln (l/\eta)$ in the dissipation range. 
Summary data for all $p$ up to $10$ are given in Fig.~\ref{fig:L_zeta_allp_fit} 
with linear extrapolation from the dissipation range to the Kolmogorov scale for the JHTDB data.

\begin{figure}
\center{\includegraphics[width=1.\textwidth]{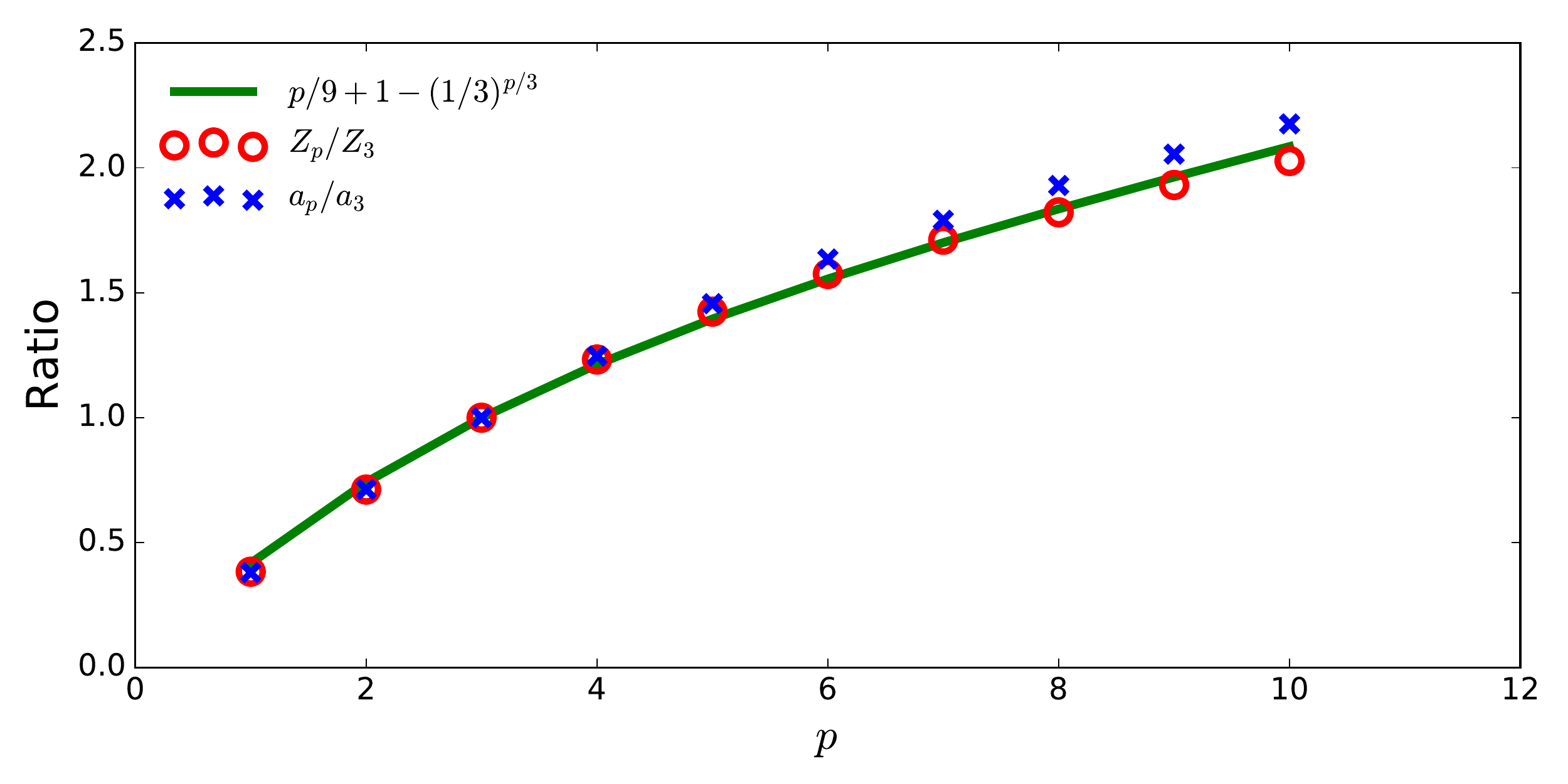}}
\caption{\label{fig:ZpZ3} The new ratio model agrees with the ratios of $Z_p/Z_3$ and of $a_p/a_3$ from JHTDB data in the dissipation range.}
\end{figure}

Recall that $\zeta_p = Z_p \ln \left(\frac{l}{\eta} \right) + a_p$ in our linear model for longitudinal velocity increment in the dissipation range. 
This equation gives a new relation for the linear slope ratio $Z_p / Z_3$ and the ratio $a_p / a_3$ as:
\begin{equation}
\label{eq:Zp_Z3}
\begin{split}
\zeta_p / \zeta_3  &=  \frac{Z_p \cdot \ln \left(\frac{l}{\eta} \right) + a_p}{ Z_3 \cdot \ln \left(\frac{l}{\eta} \right) + a_3} \\
				&=\frac{Z_p  + a_p / \ln \left(\frac{l}{\eta} \right)}{ Z_3 + a_3 / \ln \left(\frac{l}{\eta} \right)}	\ .
\end{split}
\end{equation}
Based on the model representation (\ref{eq:zetap_zeta3}) and on eq.~(\ref{eq:Zp_Z3}), we develop a model for the
longitudinal velocity structure in the dissipation range as:
\label{eq:ZpZ3_apa3}
\begin{equation}
\begin{aligned}
Z_p / Z_3 &= p/9 +1 - (1/3)^{\frac{p}{3}} \ , \\ 
a_p / a_3 &= p/9 +1 - (1/3)^{\frac{p}{3}} \ .
\end{aligned}
\end{equation}
This law is valid for the JHTDB data up to $p=10$ as shown in Fig.~\ref{fig:ZpZ3}. 

\begin{figure}
\center{\includegraphics[width=1.\textwidth]{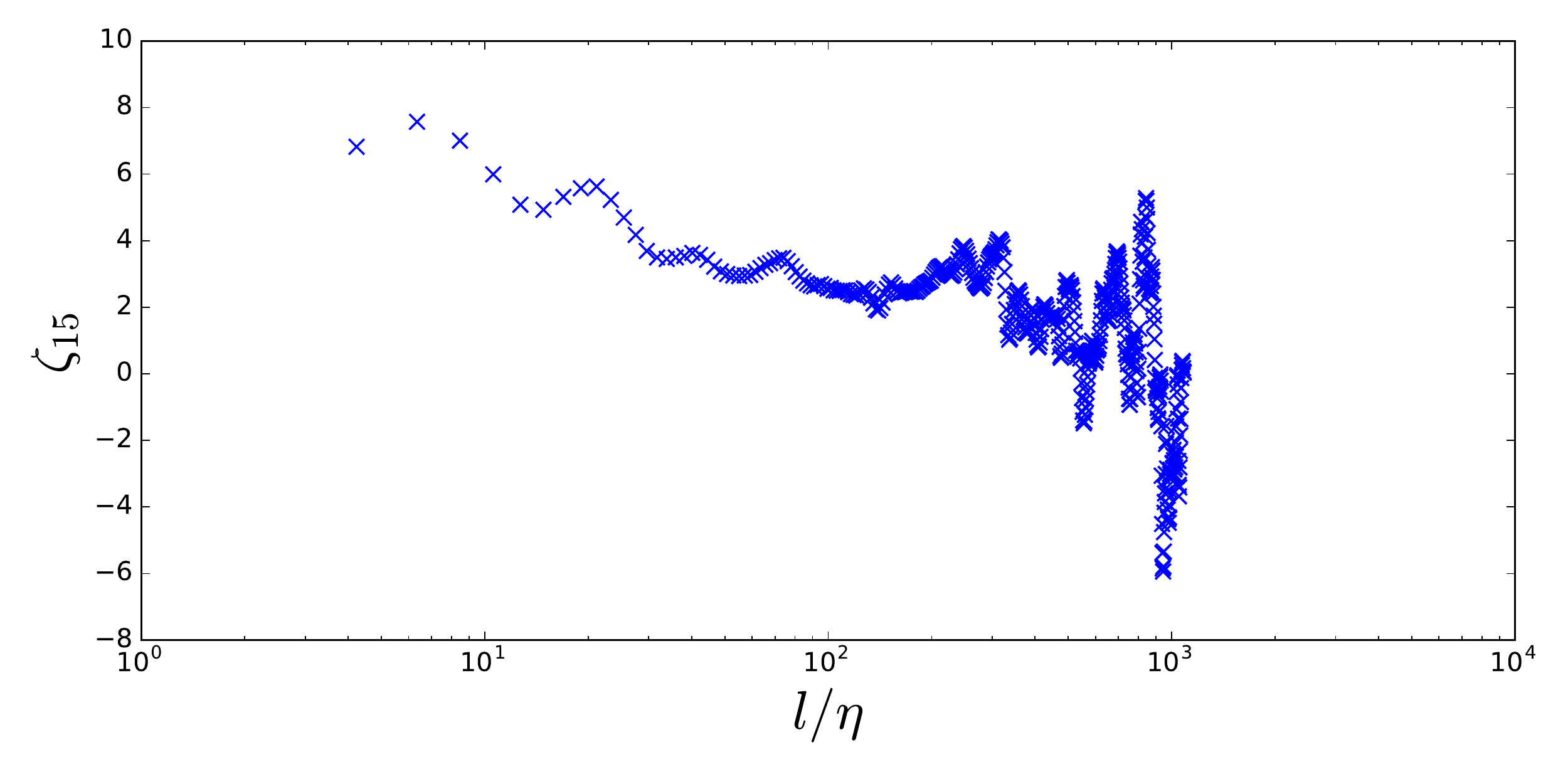}}
\caption{\label{fig:zeta_15}$\zeta_p$ is no longer linear in the dissipative range for $p>10$.}
\end{figure}

Given $Z_3$, $Z_p$ can be modeled in the dissipation range by eq.~(\ref{eq:ZpZ3_apa3}).
Fig.~\ref{fig:zeta_15} shows $\zeta_{15}$ is not linear in $\ln (l/\eta)$ in the dissipation range. This is typical for $p > 10$, and we do not include a model for $\zeta_p$ for larger $p$.
The mismatch between the data and model for larger $p$ raises the question whether the model needs to be improved, the numerical methods are unstable for high moments, or the data need improvement.

\subsection{Full \text{\large$\zeta_p$} parameterization}
\label{subsec:zeta_full}
We have developed a model that captures the longitudinal velocity increments for all of the small length scales up to and including the inertial range. The $\zeta_p$ parameterization starts with $\zeta_p = a_p$ at the Kolmogorov length scale ($l/\eta=1$) as shown in eq.~(\ref{eq:zeta_linear}).
Based on eq.~(\ref{eq:ZpZ3_apa3}), a model for $\zeta_p$ that captures the linear segment in the dissipative range for each $p$ with input $Z_3$ and $a_3$ value dependence is defined as:
\begin{equation}
\label{eq:full_parm_zetap}
\begin{split}
\zeta^{dr}_p &= Z_p  \cdot  \ln \left(\frac{l}{\eta} \right) + a_p \\
	&= \left(p/9 +1 - (1/3)^{\frac{p}{3}} \right)  \cdot \left( Z_3  \cdot  \ln \left(\frac{l}{\eta} \right) + a_3 \right) \ .
\end{split}
\end{equation}
We have parameterized $\zeta_p$ for longitudinal velocity increments $\langle |\delta_l u|^p \rangle$ at the dissipation range with a finite number of parameters for $p$ up to $10$. It is $p$ dependent, with the variables $Z_3$ and $a_3$ calculated from $\zeta_3$. 

The dissipation range begins at the Kolmogorov length scale and ends at the intersection with the theoretical value from SL, which marks the start of the inertial range. This transition occurs approximately at the Taylor micro scale. The transition point can be found by 
\begin{equation}
\left(p/9 +1 - (1/3)^{\frac{p}{3}} \right)  \cdot \left( Z_3  \cdot  \ln \left(\frac{l}{\eta} \right) + a_3 \right) = \zeta^{dr}_p = \zeta^{SL}_p = p/9 + 2 \cdot \left[1 - (2/3)^{\frac{p}{3}} \right] \ .  
\end{equation}
We solve this equation for length scale $l$ with the known variable $Z_3$ and $a_3$ from $\zeta_3$ data along with the known $\eta$.
The $p$ dependent transition points of $\zeta_p$ for $p=2, 3, 6$ are shown as the thick plus symbols in Fig.~\ref{fig:zeta236_fit}.

\section{The SL conjecture for the laminar limit}
\label{sec:sl_conject}
SL discusses the large $p$ asymptotes of $\epsilon_l^p$ in developing parameters for their $\tau_p$ inertial range methodology. They predicted dominance by vortices in this regime. We come to the same conclusion in the large $p$ asymptotes more directly through an analysis of $\zeta_p$, which reflects the intensity of the vortical structures.
We find in Fig.~\ref{fig:L_zeta_allp_fit} and eq.~(\ref{eq:full_parm_zetap}) that $\zeta_p$ is increasing in $p$ as the length $l/\eta$ moves toward the dissipation range. 

We interpret the length scale $l/\eta$ limit in terms of Taylor-Green vortices continued past the instability point. Assuming the CC model remains applicable in this range, Fig.~\ref{fig:tau2_CCp} analyzes the $CC_\infty$ model and shows a nonzero residual value for $T_2$ and similarly for all $T_p$. Line vortices do not dissipate energy and are described by the vanishing $\tau_p$ for all $p$. The presence of non-zero $T_p$ for all $p$ indicates that the line vortices occur in an unstable state, 
as occurs in a Taylor-Green vortex, continued past its singular value.

\section{Conclusions}
\label{sec:conc}
We have extended scaling laws for longitudinal velocity increments 
and energy dissipation rate structure functions 
from the inertial range to all length
scales by modifying their exponential scaling exponents.
We verify the complete parameterization models for $\tau_p$ and $\zeta_p$ in the dissipation range, i.e. from the Kolmogorov scale to the Taylor scale, limited to $p \leq 10$ for $\zeta_p$.
Our major model feature, linearity of the log scale
dissipation processes, is verified through comparison to the JHTDB and UMA data.

In local regions, even within a fully developed turbulent flow, the turbulence is not isotropic nor scale invariant due to the influence of larger turbulent structures (or their absence). For this reason, turbulence that is not fully developed is an important issue which the present analysis addresses. 

In Kolmogorov theory and in advanced multifractal scaling law
theories, $\zeta_2$ has a range in which it is a flat line of constant value in log length scale,
as observed. It has been noted  that $\tau_p$ is not
constant in the inertial range that is defined by $\zeta_2$. 
Our data analysis and data of others show no clearly defined inertial range for the energy dissipation rate.
The $\tau_p$ peak is $p$ dependent, but occurs approximately at the Taylor micro-scale.
The transition from the dissipation range to the inertial range takes place near the Taylor micro-scale.

We find that the Chen and Cao model for $\tau_p$ can be extended across all length scales for small $p$ moments. 
Our refined $CC_q$ model describes the relation between any $\tau_p$ and $\tau_q$ exponents.
The $CC_\infty$ model complements the SL analysis of vortices in the $l/\eta \to 1$ limit are given.  

\section{Declaration of Interests}
The authors report no conflict of interest.

\bibliographystyle{jfm}
\bibliography{scales}

\begin{thebibliography}{19}
\expandafter\ifx\csname natexlab\endcsname\relax\def\natexlab#1{#1}\fi
\def\au#1{#1} \def\ed#1{#1} \def\yr#1{#1}\def\at#1{#1}\def\jt#1{\textit{#1}}
  \def\bt#1{#1}\def\bvol#1{\textbf{#1}} \def\vol#1{#1} \def\pg#1{#1}
  \def\publ#1{#1}\def\arxiv#1{#1}\def\org#1{#1}\def\st#1{\textit{#1}}

\bibitem[Almalkie \& De~Bruyn~Kops(2012)]{AlmSte12}
{\sc \au{Almalkie, S.} \& \au{De~Bruyn~Kops, S.}} \yr{2012}  \at{Energy
  dissipation rate surrogates in incompressible navier–stokes turbulence}.
  \jt{Journal of Fluid Mechanics} .

\bibitem[Boldyrev {\em et~al.\/}(2002)Boldyrev, Nordlund \& Padoan]{BolNor02}
{\sc \au{Boldyrev, Stanislav}, \au{Nordlund, Ake} \& \au{Padoan, Paolo}}
  \yr{2002}  \at{Scaling relations of supersonic turbulence in star-forming
  molecular clouds}.  \jt{The Astrophysical Journal}  \bvol{573}~(2),
  \pg{678--684}.

\bibitem[Cao {\em et~al.\/}(1996)Cao, Chen \& She]{CaoCheShe96}
{\sc \au{Cao, Nianzheng}, \au{Chen, Shiyi} \& \au{She, Zhen-Su}} \yr{1996}
  \at{Scalings and relative scalings in the { {N}avier}-{{S}tokes} turbulence}.
   \jt{Phys. Rev. Lett.}  \bvol{76},  \pg{3711--3714}.

\bibitem[Chavarria {\em et~al.\/}(1995{\natexlab{{\em a\/}}})Chavarria, Baudet,
  Benzi \& Ciliberto]{ChaBau95_2}
{\sc \au{Chavarria, Gerardo}, \au{Baudet, Christophe}, \au{Benzi, R.} \&
  \au{Ciliberto, Sergio}} \yr{1995{\natexlab{{\em a\/}}}}  \at{Hierarchy of the
  velocity structure functions in fully developed turbulence}  \bvol{5}.

\bibitem[Chavarria {\em et~al.\/}(1996)Chavarria, Baudet, Benzi \&
  Ciliberto]{ChaBau96}
{\sc \au{Chavarria, Gerardo}, \au{Baudet, Christophe}, \au{Benzi, R.} \&
  \au{Ciliberto, Sergio}} \yr{1996}  \at{Scaling laws and dissipation scale of
  a passive scalar in fully developed turbulence}.  \jt{Physica D: Nonlinear
  Phenomena}  \bvol{99}~(2),  \pg{369 -- 380}.

\bibitem[Chavarria {\em et~al.\/}(1995{\natexlab{{\em b\/}}})Chavarria, Baudet
  \& Ciliberto]{ChaBau95}
{\sc \au{Chavarria, G.~Ruiz}, \au{Baudet, C.} \& \au{Ciliberto, S.}}
  \yr{1995{\natexlab{{\em b\/}}}}  \at{Hierarchy of the energy dissipation
  moments in fully developed turbulence}.  \jt{Phys. Rev. Lett.}  \bvol{74},
  \pg{1986--1989}.

\bibitem[Chen \& Cao(1995)]{CheCao95}
{\sc \au{Chen, Shiyi} \& \au{Cao, Nianzheng}} \yr{1995}  \at{Inertial range
  scaling in turbulence}.  \jt{Phys. Rev. E}  \bvol{72:R5757-R5759}.

\bibitem[Frick {\em et~al.\/}(1995)Frick, Dubrulle \& Babiano]{FriDub95}
{\sc \au{Frick, P.}, \au{Dubrulle, B.} \& \au{Babiano, A.}} \yr{1995}
  \at{Scaling properties of a class of shell models}.  \jt{Phys. Rev. E}
  \bvol{51},  \pg{5582--5593}.

\bibitem[Frisch(1996)]{Fri95}
{\sc \au{Frisch, U.}} \yr{1996} {\em Turbulence: The Legacy of {A}. {N}.
  {K}olmogorov\/}.  \publ{Cambridge: Cambridge Univeristy Press}.

\bibitem[He {\em et~al.\/}(1999)He, Doolen \& Chen]{HeChe99}
{\sc \au{He, Guowei}, \au{Doolen, Gary~D.} \& \au{Chen, Shiyi}} \yr{1999}
  \at{Calculations of longitudinal and transverse velocity structure functions
  using a vortex model of isotropic turbulence}.  \jt{Physics of Fluids}
  \bvol{11}~(12),  \pg{3743--3748}.

\bibitem[Kolmogorov(1941)]{Kol41}
{\sc \au{Kolmogorov, A.~N.}} \yr{1941}  \at{Local structure of turbulence in
  incompressible viscous fluid for very large {R}eynolds number}.  \jt{Doklady
  Akad. Nauk. SSSR}  \bvol{30},  \pg{299--3031}.

\bibitem[Kolmogorov(1962)]{Kol62}
{\sc \au{Kolmogorov, A.~N.}} \yr{1962}  \at{A refinement of previous hypotheses
  concerning the local structure of turbulence in a viscous incompressible
  fluid at high reynolds number}.  \jt{J. Fluid Mechanics}  \bvol{13},
  \pg{82--85}.

\bibitem[L.~Miller \& E.~Dimotakis(1991)]{MilDim91}
{\sc \au{L.~Miller, Paul} \& \au{E.~Dimotakis, Paul}} \yr{1991}  \at{Stochastic
  geometric properties of scalar interfaces in turbulent jets}.  \jt{Physics of
  Fluids A Fluid Dynamics} .

\bibitem[Li {\em et~al.\/}(2008)Li, Perlman, Wan, Yang, Burns, Meneveau, Burns,
  Chen, Szalay \& Eyink]{LiPerWan08}
{\sc \au{Li, Y.}, \au{Perlman, E.}, \au{Wan, M.}, \au{Yang, Y.}, \au{Burns,
  R.}, \au{Meneveau, C.}, \au{Burns, R.}, \au{Chen, S.}, \au{Szalay, A.} \&
  \au{Eyink, G.}} \yr{2008}  \at{A public turbulence database cluster and
  applications to study {L}agrangian evolution of velocity increments in
  turbulence}.  \jt{Journal of Turbulence}  \bvol{9}~(31).

\bibitem[M\"uller \& Biskamp(2000)]{MulBis00}
{\sc \au{M\"uller, Wolf-Christian} \& \au{Biskamp, Dieter}} \yr{2000}
  \at{Scaling properties of three-dimensional magnetohydrodynamic turbulence}.
  \jt{Phys. Rev. Lett.}  \bvol{84},  \pg{475--478}.

\bibitem[Novikov(1994)]{Nov94}
{\sc \au{Novikov, E.~A.}} \yr{1994}  \at{Infinitely divisible distributions in
  turbulence}.  \jt{Phys. Rev. E}  \bvol{50},  \pg{R3303--R3305}.

\bibitem[Perlman {\em et~al.\/}(2007)Perlman, Burns, Li \&
  Meneveau]{PerBurLi07}
{\sc \au{Perlman, E.}, \au{Burns, R.}, \au{Li, Y.} \& \au{Meneveau, C.}}
  \yr{2007} Data {E}xploration of {T}urbulence {S}imulations using a {D}atabase
  {C}luster.  \bt{In {\em SC '07 Proceedings of the 2007 ACM/IEEE conference on
  Supercomputing\/}}.

\bibitem[She \& Leveque(1994)]{SheLev94}
{\sc \au{She, Z.~S.} \& \au{Leveque, E.}} \yr{1994}  \at{Universal scaling laws
  in fully developed turbulence}.  \jt{Phys. Rev. Lett.}  \bvol{72},
  \pg{336--339}.

\bibitem[Zou {\em et~al.\/}(2003)Zou, Zhu, Zhou \& She]{ZouZhu03}
{\sc \au{Zou, Zhengping}, \au{Zhu, Yuanjie}, \au{Zhou, Mingde} \& \au{She,
  Zhen-Su}} \yr{2003}  \at{Hierarchical structures in a turbulent pipe flow}.
  \jt{Fluid Dynamics Research}  \bvol{33}~(5),  \pg{493 -- 508}.

\end{thebibliography}

\end{document}